\documentclass{emulateapj}
\usepackage{graphicx}
\usepackage[usenames, dvipsnames]{color}

\newcommand{\beq}	{\begin{equation}}
\newcommand{\eeq}	{\end{equation}}
\newcommand{\beqa}{\begin{eqnarray}}
\newcommand{\eeqa}{\end{eqnarray}}

\newcommand{\eee}	{$^{-3}$}
\def\simlt{\lower.5ex\hbox{$\; \buildrel < \over \sim \;$}}
\def\simgt{\lower.5ex\hbox{$\;. \buildrel > \over \sim \;$}}

\font\tenbi=cmmib10 
\newfam\bifam  \textfont\bifam=\tenbi

\font\tenbr=cmbx10
\newfam\brfam  \textfont\brfam=\tenbr
\newcommand{\p}{\partial}

\font\squinttenbi=cmbx10 at 9pt
\scriptfont\brfam=\squinttenbi

\def\vecnabla{
              \setbox1=\hbox{$\bigtriangledown$}
                           \raise.45ex\hbox{$\bigtriangledown$\hskip-.97\wd1
                           $\bigtriangledown$\hskip-.97\wd1
                           $\bigtriangledown$\hskip-.97\wd1}
                           \raise.47ex\hbox{$\bigtriangledown$}}

\def\rsun{\ifmmode {\rm R}_{\mathord\odot}\else $R_{\mathord\odot}$\fi}
\def\msun{\ifmmode {\rm M}_{\mathord\odot}\else $M_{\mathord\odot}$\fi}
\def\lsun{\ifmmode {\rm L}_{\mathord\odot}\else $L_{\mathord\odot}$\fi}

\newcommand{\mup}	{\mu_{\Phi}}
\newcommand{\mupeff}	{\mu_{\Phi, {\rm eff}}}

\newcommand{\kms}	{{\rm km}\, {\rm s}^{-1}}

\def\tmb{\ifmmode {T_{\rm mb}^{13}(x,y,v)}\else $T_{\rm mb}^{13}(x,y,v)$\fi}

\begin{document}

\title{Impact of Protostellar Outflows on Turbulence and Star Formation Efficiency in Magnetized Dense Cores}

\author{Stella S.~R.~Offner$^1$ $^2$}
\author{Jonah Chaban$^1$}
\affil{$^1$Department of Astronomy, University of Massachusetts, Amherst, MA 01003}
\affil{$^2$Department of Astronomy, University of Texas, Austin, TX}
\email{soffner@astro.umass.edu }

\begin{abstract}
The star-forming efficiency of dense gas is thought to be set within cores by outflow and radiative feedback. We use magneto-hydrodynamic simulations to investigate the relation between protostellar outflow evolution, turbulence and star formation efficiency. We model the collapse and evolution of isolated dense cores for $\gtrsim$ 0.5 Myr including the effects of turbulence, radiation transfer, and both radiation and outflow feedback from forming protostars. We show that outflows drive and maintain turbulence in the core environment even with strong initial fields. The star-formation efficiency decreases with increasing field strength, and the final efficiencies are $15-40$\%. The Stage 0 lifetime, during which the protostellar mass is less than the dense envelope, increases proportionally with the initial magnetic field strength and ranges from $\sim 0.1-0.4$ Myr. The average accretion rate is well-represented by a tapered turbulent core model, which is a function of the final protostellar mass and is independent of the magnetic field strength. By tagging material launched in the outflow, we demonstrate that the outflow entrains about $3$ times the actual launched gas mass, a ratio that remains roughly constant in time regardless of the initial magnetic field strength. However, turbulent driving increases for stronger fields since momentum is more efficiently imparted to non-outflow material.  The protostellar outflow momentum is highest during the first 0.1 Myr and declines thereafter by a factor of $\gtrsim 10$ as the accretion rate diminishes.
\end{abstract}

\section{Introduction}

Forming stars impact their natal environment through several ``feedback" processes including radiative heating and bipolar mass outflows. 
The interaction between protostars and their environment shapes fundamental astrophysical quantities such as the distribution of stellar masses, the properties of protostellar cores, and the longevity and energetics of molecular clouds \citep{Offner14ppvi,krumholz14ppvi}.  

Outflowing material is characterized by two observable components: a highly collimated jet mainly comprised of atomic gas and a broader, colder outflow made primarily of molecular gas \citep{frankppvi14}. Jets from low-mass stars achieve velocities of $\sim 100~\kms$, while molecular outflows exhibit velocities of $\sim 1-10~\kms$. Each can span up to several parsecs in length.

Protostellar outflows are thought to arise from the winding of the magnetic field lines that thread the inner accretion region. Coupling between the field and gas causes some incoming material to be flung outwards above and below the accretion disk. The high-velocity protostellar jet is thought to be produced by gas ejected close the stellar surface \citep{shu94}, while the wider angle molecular outflow may be composed of material from the inner accretion disk \citep{pelletier92}. However, the details of the engine responsible for launching protostellar outflows remain uncertain, and it is debated whether outflow and jet signatures correspond to separate physical mechanisms.  

Since outflows are launched during the embedded phase on sub-AU scales, the launching region is both deeply obscured and challenging to resolve. Jet and outflow morphologies on parsec scales have been studied for decades; the first jets were discovered by Herbig and Haro independently in the 1950s \citep{herbig50,haro52}, although their collimated nature was not recognized for many years. On smaller scales, the complex interactions between the outflow, natal core and cloud are difficult to disentangle. Velocity information is only available along one sightline and motions of a few $\kms$ or less are often lost in the general cloud turbulence \citep[e.g.,][]{arce01}. Consequently, the youngest and lowest mass outflows are especially challenging to identify and study. 

Protostellar outflows broadly influence the star formation process in a variety of ways. As the outflow exits the natal dense core, it sweeps up and removes dense material, such that much of the molecular component is most likely entrained dense gas. This effectively reduces the amount of material available for accretion and may contribute to the process of core dispersal that ultimately terminates accretion \citep{dunham14,zhang16}. Outflow entrainment may also be responsible for the apparent offset in peak mass between the stellar initial mass function (IMF) and the mass function of dense cores \citep{Offner14ppvi}.  Outflows in aggregate can replenish turbulent motions and may impact the global cloud evolution on scales of a few parsecs \citep{matzner00,swift08,nakamura11,plunkett13}.

Numerical simulations have provided a complementary avenue to explore the physics and impact of protostellar outflows. Simple outflow launching models have demonstrated that the interactions of outflows in low-mass forming clusters can sustain turbulence and slow star formation \citep{nakamura07,wang10,carroll10,hansen12,federrath14,li17}. The addition of outflows to cluster simulations reduces the star formation efficiency by $\sim 30$\% \citep{hansen12,myers14,federrath14}. Studies of protostar formation in isolated cores find that outflows reduce the efficiency of dense gas by $\sim 30-40$\% \citep[][henceforth, OA14]{machida13,Offner14b}. 
Outflows may also efficiently drive turbulence within the core over the main accretion phase (OA14).  

Despite these advances, numerical simulations of protostellar outflows remain challenging due to the complicated physics and computational expense involved.
In current state-of-the art calculations, the protostellar jet can be launched self-consistently and resolved down to $\sim R_\odot$ scales \citep{tomida14,tomida15}. However, such calculations can follow only a very limited evolutionary time ($\sim \mathcal{O}$(yr)). Other groups modeling outflows that aim to address time evolution follow one of two approaches.  In the first approach, calculations include magnetic fields, allowing the outflow to launch self-consistently, but they do not resolve sub-AU scales. This produces less collimated outflows that are limited to maximum velocities of a few tens of $\kms$ \citep{seifried12,machida13,lewis17}.  Alternatively, calculations include a sub-grid model that parameterizes outflow launching and removes most of the resolution dependence \citep{nakamura07,wang10,cunningham11,peters14,federrath14}. These simulations may also include magnetic fields but do not rely on the fields to produce outflow behavior. It is worth noting that in both cases, non-ideal MHD effects, which are demonstrably important for accretion disk formation and evolution, are often incomplete or absent \citep{lippvi14}.

In this work, we extend the study by OA14 to investigate the impact of outflows on turbulence and star formation efficiency in magnetized, turbulent cores with radiation feedback. We adopt a sub-grid model for the outflow launching, calibrated using observations, that allows us to follow the collapse and protostellar evolution for $\gtrsim$0.5 Myr. For our calculations, we also adopt an updated version of {\sc Orion}, which includes a ``tracer field", permitting us to tag and track outflow material. We describe our methods in \S\ref{methods} and present results in \S\ref{results}, which we separate into sub-sections devoted to the gas (\S\ref{gasprop}), protostar (\S\ref{protoprop}) and outflow evolution (\S\ref{outflowprop}). 
We summarize our conclusions in \S\ref{conclude}.




\section{Methods}\label{methods}

\subsection{Physical Parameters and Scalings}

In this study we explore the relation between the initial core mass, velocity dispersion and magnetic field strength on star-formation efficiency.  These parameters are broadly responsible for setting the collapse time, binding energy and stellar mass. Thus, we re-frame these quantities in terms of physical parameters describing the relative importance of gravity, magnetic fields and thermal pressure. 

The magnetic flux through a core with radius $R_{\rm core}$ and uniform magnetic field $B_0$ is 
$\Phi = \pi R_{\rm core}^2B_0$ \citep{mouschovias76}.
For a core of mass $M_{\rm core}$ the mass-to-flux ratio is 
$\mup = M_{\rm core} / M_{\Phi},$ 
where 
\beq
M_{\Phi} \simeq \frac{\Phi}{2\pi G^{1/2}}
\eeq
is the magnetic critical mass. Cores with $\mup > 1$ are ``super-critical" and unstable to gravitational collapse. Cores with $\mup < 1$ are ``sub-critical'' and are magnetically supported. Observed cores appear to have $\mup\simeq 2-3$, although this is uncertain to within a factor of two \citep{crutcher09, crutcher12}. Normalizing to 
a set of fiducial core properties, we have
\beqa
M_{\Phi} &\simeq& 2.0 \left(  \frac{R_{\rm core}}{0.065\, {\rm pc} }  \right)^2 \left(  \frac{B_0} {51.5\, \mu {\rm G}}  \right) \msun \\
\mup  &=& 2 \left(  \frac{M_{\rm core}}{4\, \msun}  \right) \left(  \frac{0.065\, {\rm pc} }{R_c}  \right)^2 \left(  \frac{51.5\, \mu {\rm G}}{B_0}  \right).
\eeqa

These expressions, however, do not account for thermal (or turbulent) pressure. \citet{mckee89} proposed a modification to the critical mass estimate: 
\beq 
M_{\rm cr}  \simeq M_{\rm BE} + M_{\Phi},
\eeq
where $M_{\rm BE}$ is the maximum stable mass (Bonnor-Ebert mass). The Bonnor-Ebert mass is given by:
\beq
M_{\rm BE} = 1.18 \frac{c_s^4}{(G^3 P_0)^{1/2}},
\eeq
where $c_s$ is the isothermal sound speed, and $P_0$ and $\rho_0$ are the surface pressure and density, respectively. For a fiducial temperature and density of $T=10\,$K and $\rho_0=2.39\times 10^{-19}$g cm$^{-3}$, $M_{\rm BE}=0.47 \msun$.


Thus, we define the effective mass-to-flux ratio:
\beq
\mupeff = \frac{M_{\rm core}}{M_{cr}}, 
\eeq
where $\mupeff \lesssim 1$ will be stable against collapse. We note that this criterion does not exactly indicate stability given our initial parameters since both the mean core temperature and root-mean-squared (rms) magnetic field increase slightly as a result of the initial turbulence. 

The relative importance of turbulence and gravity is given by the virial parameter \cite[e.g.,][]{bertoldi92}:
\beq
\alpha = 5 \sigma^2 R_c/ (GM_{\rm core}),
\eeq
where $\sigma$ is the 1D velocity dispersion. This implicitly assumes that turbulence acts as an isotropic pressure. Cores with $\alpha \gtrsim 2$ are considered to be gravitationally unbound and will not collapse.

\begin{deluxetable*}{lccccccccc}
\tablecolumns{10}
\renewcommand{\tabcolsep}{0.07cm}
\tablecaption{Model Properties \label{simprop}}
\tablehead{ \colhead{Model\tablenotemark{a}} &  
  \colhead{$M_{\rm core}$($\msun$) } &
   \colhead{$l_{\rm max}$} & 
  \colhead{$\Delta x_{\rm max}$(AU)} &
 \colhead{$B_z$($\mu$G)}&
 \colhead{$M_{\phi}$($\msun$)}&
  \colhead{$\mu_{\phi}$}&
  \colhead{$M_{\rm cr,eff}$($\msun$)}&
 \colhead{$\mupeff$} &
 \colhead{$v_i$(kms$^{-1}$) } }
\startdata
M4P1.5\tablenotemark{b}     & 4.0 & 5 & 26& 68.67 & 2.67 & 1.5&3.14 & 1.28 & 0.52   \\
M4P2.5\tablenotemark{b}      & 4.0 & 5 & 26   & 41.20 & 1.6 &2.5 &  2.07 & 1.93 & 0.52    \\ 
M4P2.5b\tablenotemark{c}     & 4.0 & 5 & 26   & 41.20 & 1.6 & 2.5 & 2.07 & 1.93 & 0.52      \\
M4P2.5l6     & 4.0 & 6 & 13   & 41.20 & 1.6 & 2.5 & 2.07 & 1.93 & 0.52      \\
M4P2.5nt     & 4.0 & 6 & 26   & 41.20 & 1.6 & 2.5 & 2.07 & 1.93 & 0.0      \\
M4P5     & 4.0 & 5 & 26   & 20.6 & 0.80 & 5 & 1.27 &3.15  &0.52     \\
M4P18     & 4.0 & 5 & 26   & 5.6 & 0.22 & 18.4 & 0.69 & 5.80 & 0.52  \\ 
M4P$\infty$\tablenotemark{d}     & 4.0 & 5 & 26   & 0 & 0 & $\infty$  & 0.47 & $\infty$ & 0.52  \\ 
M8P2.5b\tablenotemark{c}      & 8.0 & 5 & 26  & 82.4 &  3.2   & 2.5 & 3.53  & 2.27   & 0.74     
\enddata
\tablenotetext{a}{The model name, gas mass, maximum AMR level, cell size on the maximum level, magnitude of the initial magnetic field in the $z$ direction, the magnetic mass, mass-to-flux ratio, effective magnetic critical mass, effective mass-to-flux ratio, and initial rms turbulent velocity.  The molecular cores all have initial radii of $R_c=$0.065 pc, a uniform density, initial temperature $T_c=$10 K, and initial virial parameter of $\alpha=$2.0.}
\tablenotetext{b}{Calculation analyzed in dust polarization study published in \citet{lee17}.}
\tablenotetext{c}{Calculation analyzed in multiplicity study carried out in \citet{Offner16}.}
\tablenotetext{d}{Non-magnetized calculation presented and analyzed in OA14 (their run th0.01).}
\end{deluxetable*}


\subsection{Initial Conditions}

Following OA14, the simulations begin with a sphere of uniform density, 10 K gas and radius $R_c = 0.065$ pc, which is pressure-confined by a warm ($10^3$~K), low-density ($\rho_c/100$) medium. The domain extent is twice the initial diameter of the dense core ($L=4R_c=0.26$ pc). We adopt outflow boundary conditions, such that unbound, high-velocity gas can leave the domain. The initial magnetic field is uniform in the $z$ direction: $\vec B = B_0 \hat z$. 

The gas velocities in the core are initialized with a turbulent random field that has a flat power spectrum over wavenumbers $k=2-4~L^{-1}$, which maps to physical scales $2R_c- R_c$. The velocity field is normalized to satisfy the specified velocity dispersion.  We solve for the evolution of these conditions according to the equations given in \S\ref{nummethods}.

All the runs, with the exception of M4P$\infty$, begin with the same turbulent random seed and have a core virial parameter of $\alpha = 2.0$. Although this corresponds to a velocity dispersion that exceeds the expected linewidth-size relation \citep[e.g.,][]{MandO07}, motions decay quickly and become sub-sonic. The protostellar outflow provides the only additional energy input.  Table \ref{simprop} summarizes the simulation properties for each model. 

We adopt a basegrid resolution of $64^3$. The dense core is refined by two AMR levels using a density threshold refinement criterion. Each subsequent AMR level increases the resolution by a factor of 2. As the gas collapses, additional fine levels are inserted when the density exceeds the Jeans condition for a Jeans number of $N_J=0.125$ \citep{truelove97}. Cells are also added in locations with strong gradients in the radiation energy density: $\Delta E_R/ E_R > 0.15$. 
For our fiducial resolution and assuming 10~K gas, a sink particle forms when the density exceeds $\sim 8 \times 10^{-15}$g~cm$^{-3}$. These particles interact with the gas, accrete, radiate and launch protostellar outflows as described in \S\ref{nummethods}.

In one set of initial conditions a second particle forms. We study the impact of binarity on the core evolution by carrying out two runs: one in which the binary merges around $200$~AU (M4P2.5) and one in which we follow the pair throughout the calculation (M4P2.5b).
 
Three of the runs in Table \ref{simprop} have been analyzed in complementary studies of protostellar multiplicity and outflow-field alignment. \citet{Offner16} explored the impact of different random seeds for three initial core masses and the subsequent formation of multiple systems. They used identical initial conditions to those here but analyzed only simulations producing multiple protostars, which included runs M4P2.5b and M8P2.5b. 
 \citet{lee17} post-processed runs M4P2.5 and M4P1.5 to produce synthetic polarization measurements, which they compared to the observed distribution of angles between outflows and magnetic fields in protostellar cores \citep{hull13}.

\subsection{Numerical Methods}\label{nummethods}

We  perform the numerical study using the magnetohydrodynamics (MHD) adaptive mesh refinement (AMR) code {\sc Orion2}  \citep{klein99,truelove98,li12}. The calculations include self-gravity and radiative transfer in the flux-limited diffusion approximation \citep{krumholzkmb07}. {\sc Orion2} evolves the magnetic field with a constrained transport MHD scheme \citep{li12}, which is based on the finite volume formalism first implemented by \citet{mignone12}.  The full magneto-hydrodynamic equations solved by {\sc orion2} for our calculations are: 
\begin{eqnarray}
\frac{\p\rho}{\p t} &=& - \nabla\cdot(\rho\mathbf{v}) - \sum_i \dot{M}_{a,i} W_a(\mathbf{x}-\mathbf{x}_i)  \nonumber \\
&& + \sum_i \dot{M}_{o,i} W_o(\mathbf{x}-\mathbf{x}_i)  \label{eqn:continuity} \\
\frac{\p (\rho\mathbf{v})}{\p t} &=& -\nabla\cdot(\rho\mathbf{v}\mathbf{v}) - \nabla \left(P + \frac{B^2}{8\pi} \right) + \frac{1}{4\pi} \mathbf{B}\cdot\nabla\mathbf{B} \nonumber \\
&& - \rho \nabla(\phi) - 
\sum_i \dot{\mathbf{p}}_{a,i} W_a(\mathbf{x}-\mathbf{x}_i) \nonumber \\ 
&& + \sum_i \dot{\mathbf{p}}_{o,i} W_o(\mathbf{x}-\mathbf{x}_i) \label{eqn:euler} \\ 
\frac{\p\rho e}{\p t} &=& -\nabla\left[ \left(\rho e+P+\frac{B^2}{8\pi}\right)\mathbf{v} - \frac{1}{4\pi} \mathbf{B}(\mathbf{v}\cdot\mathbf{B}) \right] \nonumber \\
&& - \rho\mathbf{v}\cdot\nabla(\phi) - \kappa_{\rm 0P}\rho(4\pi B-c E_{\rm R}) \nonumber \\ 
&& 
- \left(\frac{\rho}{m_{\rm p}}\right)^2\Lambda(T_{\rm g})   
\nonumber \\
&& - \dot{\mathcal{E}}_{a,i} W(\mathbf{x}-\mathbf{x}_i) + \dot{\mathcal{E}}_{o,i} W(\mathbf{x}-\mathbf{x}_i) \label{eqn:energy} \\
\frac{\p E_{\rm R}}{\p t} &=& \nabla\cdot\left(\frac{c\lambda}{\kappa_{\rm 0R}\rho}\nabla E_{\rm R}\right) + \kappa_{\rm 0P}\rho (4\pi B_{\rm P}-c E_{\rm R}) \nonumber \\
&& + \left(\frac{\rho}{m_{\rm p}}\right)^2\Lambda(T_{\rm g}) + \sum_i \dot{L}_{i} W(\mathbf{x}-\mathbf{x}_i) \label{eqn:radenergy} \\
&&\frac{d}{dt}M_i = \dot M_i \\
&& \frac{d}{dt}\mathbf{x}_i = \frac{\mathbf{p}_i }{M_i}\\
&& \frac{d}{dt}\mathbf{p}_i = -{M_i}\nabla \phi + \dot \mathbf{p}_i \\
\frac{\p \mathbf{B}}{\p t} &=& \nabla \times (\mathbf{v}\times\mathbf{B}) \label{eqn:induction}
\end{eqnarray}
Equations \ref{eqn:continuity}-\ref{eqn:energy} describe conservation of mass density ($\rho$), momentum density ($\rho \mathbf{v}$) and energy density ($e$), where $P$, $\mathbf{v}$, $\mathbf{B}$ and $\phi$ are the gas pressure, velocity, magnetic field and gravitational potential, respectively.  $\Lambda(T_{\rm g})$ describes the cooling rate by atomic lines and continuum, a term that becomes significant only when $T\gtrsim 10^3$~K and dust sublimes \citep[e.g.,][]{cunningham11}. $\dot M_i$, $\dot{\mathbf{p}}_i$ and $\dot{\mathcal{E}}_i$ are the rates forming stars exchange mass, momentum and energy with the gas; the subscripts $a$ and $o$ denote accretion and outflow, respectively.
The summations occur over all particles. 
The $\mathbf{x_i}$ represent the particle locations, where the mass accretion and outflow regions span four and eight, cells, respectively, centered on the sink particle. Mass, momentum, and energy are removed or deposited into these cells according to a weighting kernel,  $W$ \citep{krumholz04,cunningham11}.

Equation \ref{eqn:radenergy} describes the evolution of the radiation energy density ($E_{\rm R}$) for gray flux-limited diffusion, where $\lambda$ is the flux-limiter 
\citep[see][for more detail]{krumholzkmb07}. 
We neglect radiation pressure, since the focus of the paper is low-mass star formation \citep[e.g.,][]{Offner09}. 
The coefficients $\kappa_{\rm 0P}$ and $\kappa_{\rm 0R}$ are the Planck and Rosseland averaged dust opacities, and $B_{\rm P}$ is the Planck function. 
For the dust opacities we adopt piecewise-linear fits to the dust model from \citet{semenov03} as given in \citet{cunningham11}. 
Equation \ref{eqn:induction} is the induction equation, which describes the time-evolution of the magnetic field in the ideal  limit, i.e., the gas and field are well-coupled by definition. In this constrained transport scheme $\nabla\cdot\mathbf{B}=0$ is enforced to machine accuracy by construction. 



 {\sc Orion2} calculates the gravitational potential of the gas 
by solving Poisson's equation: 
\begin{equation}
\nabla^2 \phi = 4\pi G \left[ \rho + \Sigma_i M_i \delta(x-x_i) \right], \label{eqn:poisson}
\end{equation}
where $\delta$ is the Dirac delta function.

Sink particles form when the Jeans condition for a Jeans number $N_J=0.25$ is exceeded on the maximum AMR level \citep{krumholz04}. In these calculations, each sink particle represents an individual forming star. The associated feedback follows a sub-grid model for protostellar evolution, including radiative heating \citep{Offner09} and protostellar outflows \citep{cunningham11}. Our evolution model takes into account the accretion history, which determines the protostellar size and nuclear state as well as the protostellar mass. The direction of angular momentum of the accreting material determines the outflow launching direction, as described in  \citet{fielding15}.  By construction, magnetic flux does not accrete onto sink particles. Physically, this treatment is supported by the observation that stellar magnetic fluxes are much smaller than the magnetic fluxes of the dense cores that form them \citep[e.g.,][]{cunningham12}.  By default, particles merge if they approach within 8 cells and if either particle mass is less than 0.1~\msun \citep{krumholz04}. This choice inhibits the formation of sink particles in  dense gas that lies within the accretion and outflow region \citep[e.g.,][]{krumholz04}.

At sufficiently high resolution, solution of these magnetohydrodynamics equations produces a self-consistently magnetically launched protostellar outflow \citep[e.g.][]{tomida14}; however, achieving the observed velocities of protostellar jets requires resolution comparable to the protostellar radius. This resolution is too computationally expensive to follow over timescales of $0.1-1$~Myr. Instead, we adopt a protostellar outflow model based on \citet{matzner00}. In this formulation, the outflows are described by a collimation angle, $\theta_c$, and wind launching fraction, $f_w$. This parameterization can represent either a disk-wind \citep{pelletier92} or X-wind \citep{shu88} model. Here, we adopt $\theta_c=$0.01 and $f_w = 0.21$, such that 21\% of the accreting material is launched in the outflow.  The launching velocity is a fixed fraction, $f_k=0.3$, of the Keplerian velocity, $v_{\rm K} = f_k \sqrt{GM_*/R_*}$ at the protostellar surface, where $M_*$ and $R_*$ are the instantaneous protostellar mass and radius, respectively.  These parameters are consistent with observations of outflow momentum as discussed by \citet{cunningham11}.  
	Our specific parameter choices represent an X-wind launching mechanism. Although we note OA14 found outflow entrainment and turbulence were qualitatively similar for different collimation angles.

In strong magnetic field calculations, fields efficiently remove angular momentum and thus no accretion disk forms at our fiducial grid resolution of $\sim26$AU. However, our calculations do include turbulence, and some magnetic reconnection occurs on the grid scale, such that the gas is not perfectly coupled to the field on small scales, a process termed ``reconnection diffusion"  \citep{lazarian12,lazarian15,li15}. This effect allows disks to form in ideal MHD calculations with turbulence and resolutions of a few AU \citep[e.g.,][]{krumholz12,myers14}. Here, without a resolved accretion disk, at a given time the protostellar mass is an upper limit.

\section{Results}\label{results}

\subsection{Gas Properties and Evolution}\label{gasprop}

\begin{figure*}
\begin{center}
\includegraphics[scale=1.6]{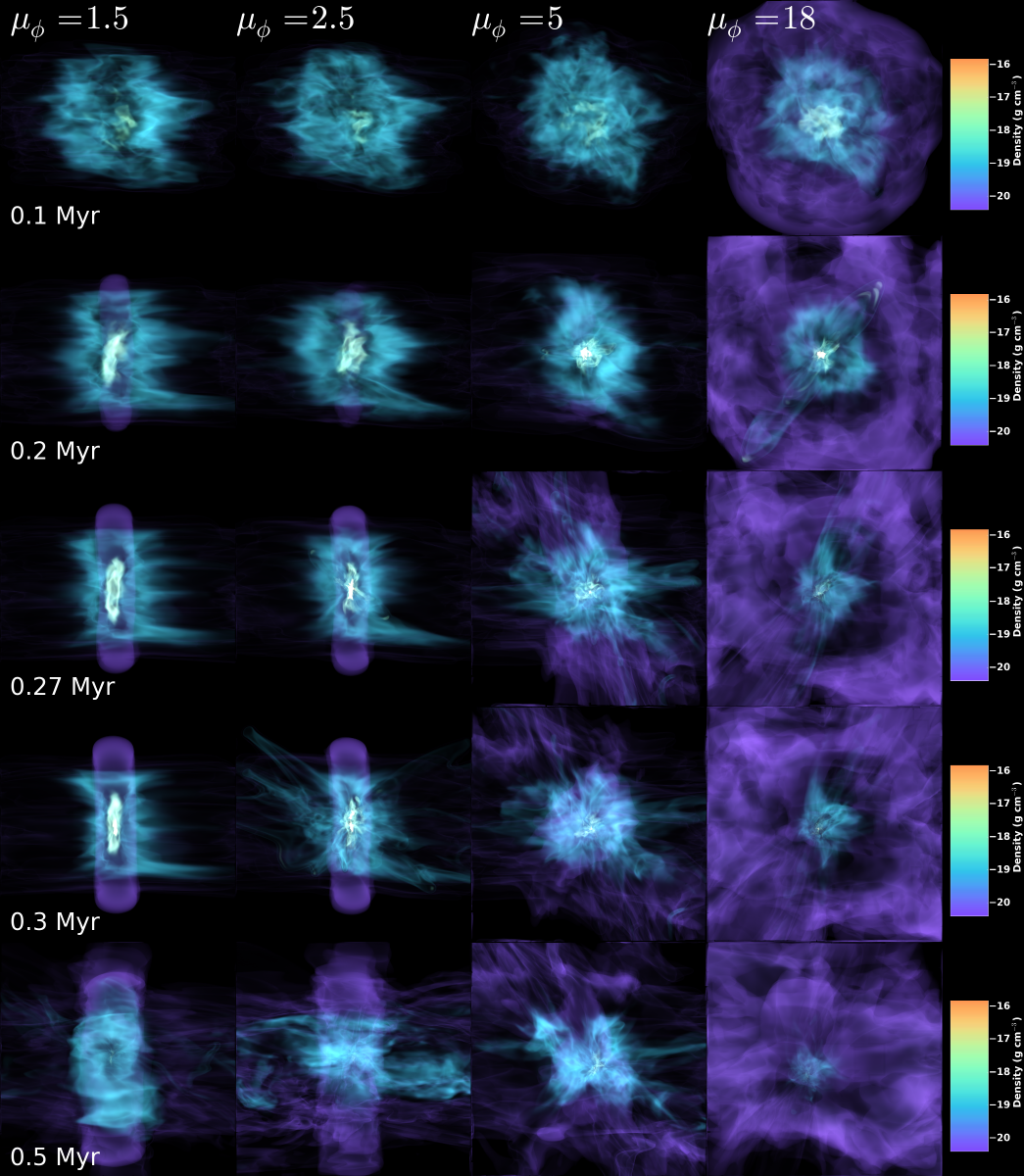}
\end{center}
\caption{Volume rendering of gas density for different models and times. The magnetic field is initially uniform in the horizontal direction.
\label{vr_Density} }
\end{figure*}

\subsubsection{Gas Morphology}\label{gasmorph}

The simulation gas morphology is strongly influenced by the initial magnetic field strength.  Figure \ref{vr_Density} shows the gas distribution as a function of time and mass-to-flux ratio.  The magnetic field, initially uniform in the $\hat z$ direction, introduces a preferred direction for collapse. The gas, which is strongly coupled to the field, collapses preferentially along the field lines, forming a dense oblate structure that appears filamentary when viewed edge-on. The lower the mass-to-flux ratio, the more quickly turbulence dissipates.  In contrast, the weaker field cases display more spherical symmetry and maintain more density substructure. In addition, they become centrally condensed more quickly, since gas can easily move perpendicular to the initial field direction. Thus, protostar formation occurs at earlier times for larger $\mu_\phi$.

Figure \ref{vr_Density} qualitatively shows that the magnetic field also influences the efficacy of feedback. In the strong field cases, the gas remains confined to the central plane, and the outflow is less efficient at dispersing the gas at any given time. Outflowing gas travels more readily along the bulk field direction.  As in OA14, the weak-field simulations show gas structure and turbulence extending well-beyond the initial core radius.

\begin{figure}
\begin{center}
\includegraphics[scale=0.46]{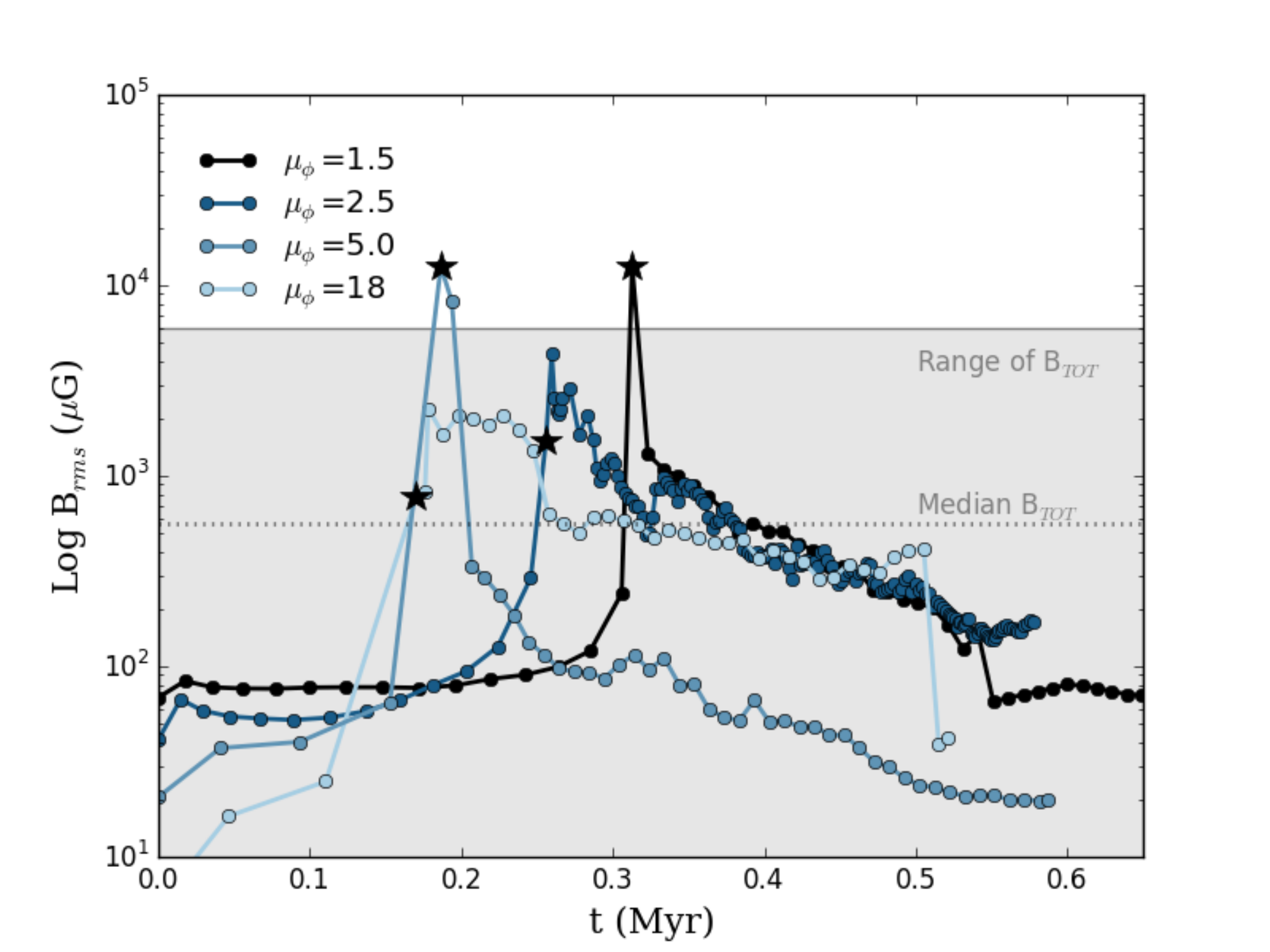}
\end{center}
\caption{Log of the mass-weighted rms magnetic field strength as a function of time.  Stars mark the time of protostar formation. The gray region indicates the range of the total magnetic field 
reported by \citet{crutcher12} for observed cores. The total field is estimated to to be twice the measured line-of-sight field ($B_{\rm TOT}=2B_{\rm los}$). The horizontal dashed black line indicates the median value of $B_{\rm TOT}$ obtained from Zeeman measurements of CN \citep{falgarone08}. \label{bvst} }
\end{figure}

\subsubsection{Magnetic Field Evolution}\label{bfield}

The global magnetic field is shaped by the initial turbulence and subsequent collapse. Figure \ref{bvst} shows the evolution of the rms magnetic field as a function of time. In the weaker field cases, the initial turbulence acts to enhance the rms field, and all the cores approach average field strengths of $\sim 50 \mu$G for times $t\leq 0.1$Myr prior to collapse.  As collapse commences, the gas becomes centrally concentrated and drags the field inwards. This coincides with a sharp increase in the global $B_{\rm rms}$ ($t=0.15-0.3$ Myr), which is concurrent with protostar formation. Cores with initially weaker field collapse on shorter timescales \citep[see also]{lewis17}.

 The degree to which the outflow drives turbulence and promotes field tangling dominates the field evolution after the protostar forms. At late times the sequence of mean field strengths is not one-to-one with the initial magnetic field, and the runs with $\mu_\Phi=18$ and $\mu_\Phi=2.5$ converge to similar values; the field of the former is enhanced due to tangling created by the outflow-core interaction, while the latter inherits the strong field from its initial conditions. These differences may also shape the magnetic field morphology on scales of $\sim$100-1,000 AU, which exhibits less order for weaker fields \citep{hull17}. 
 
\begin{figure}
\begin{center}
\includegraphics[scale=0.46]{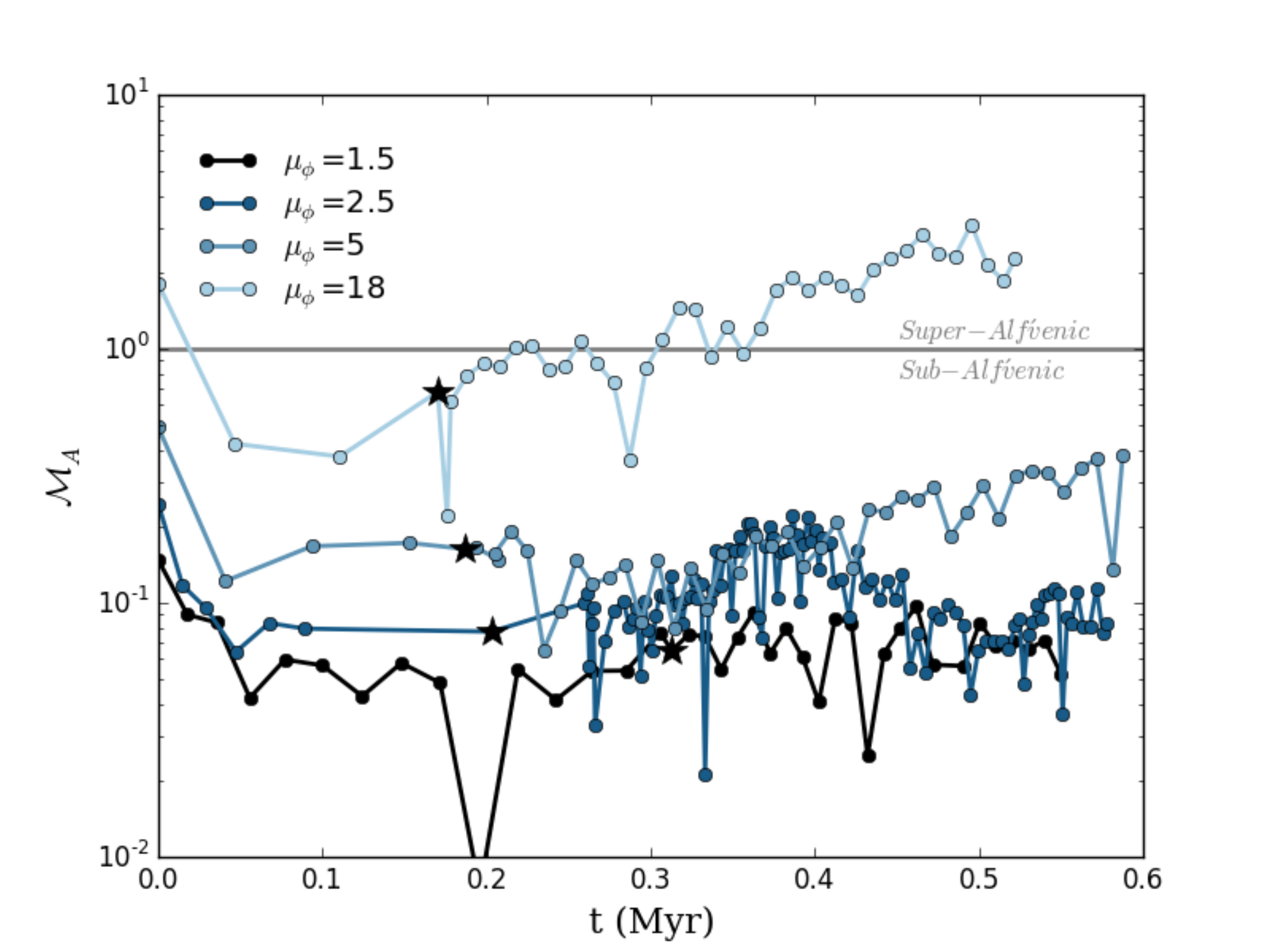}
\end{center}
\caption{ Mass-weighted Alfv\'en Mach number as a function of time for gas with $n_{\rm H} \ge 10^4$ cm$^{-3}$. Black stars mark the time of protostar formation. \label{mavst} }
\end{figure}
 
All the calculations have mean magnetic fields that are consistent with the gray shaded region in Figure \ref{bvst}. This region marks the magnetic field obtained from Zeeman measurements of OH and CN \citep{crutcher12}.  There are brief periods when the field exceeds the measurements, but these values are dominated by the field on small scales, which would not be resolved by the observational beam. The dotted line indicates the median magnetic field obtained from Zeeman observations of dense gas probed by CN  \citep{falgarone08}. The gas in the \citet{falgarone08} survey is estimated to have a mean H$_2$ density of $4.5 \times 10^5$ cm$^{-3}$. The simulated $4\msun$ cores have an initial, uniform mean density of $5.1 \times 10^4$ cm$^{-3}$, and they show good agreement with the median field value following collapse.


The relative importance of the magnetic field and turbulence is described by the Alf\'ven Mach number:
\beq 
\mathcal{M}_{\rm A} = \sqrt{ \frac{4 \pi \rho \mathbf{v}^2}{\mathbf{B}^2} }
\eeq 
Figure \ref{mavst} shows the Alf\'ven Mach number as a function of time for the different magnetic field strengths. The weakest field run, M4P18, is  super-Alf\'venic for most of the evolution, becoming sub-Alf\'venic only during the collapse phase when the initial turbulence decays and the field is enhanced by gravitational contraction. The remainder of the models are sub-Alf\'venic at all times. Runs M4P5 and M4P2.5 have comparable Alf\'venic Mach numbers for $\sim$0.15 Myr after sink formation, which is caused by enhancement of the magnetic field in M4P5 as a result of the outflow-core interaction. The magnetic similarity between these runs, which coincides with the main accretion phase, produces very similar protostellar evolution as discussed in \S\ref{protoprop}.





\begin{figure}
\begin{center}
\includegraphics[scale=0.46]{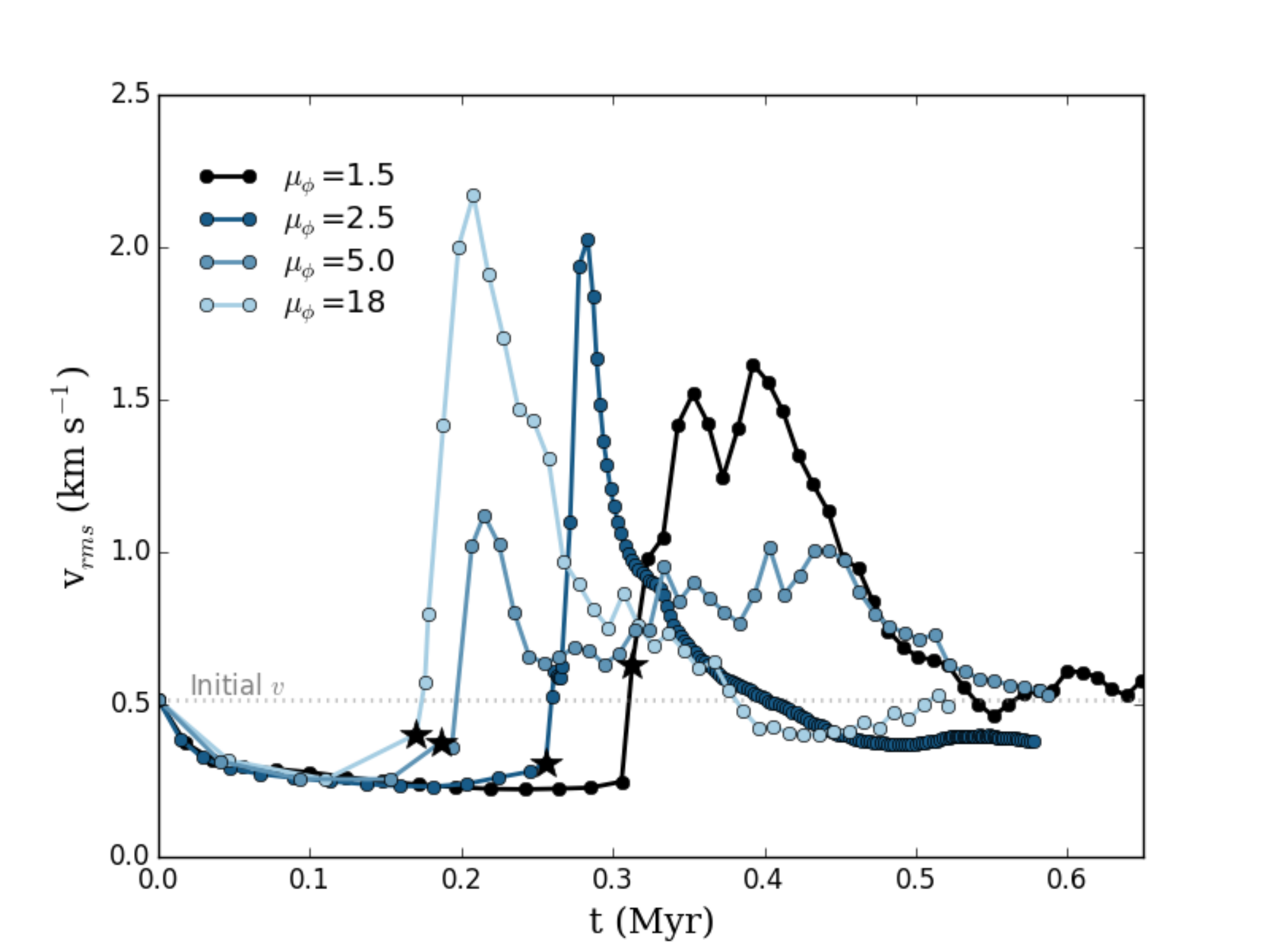}
\hspace{0.1in}
\includegraphics[scale=0.46]{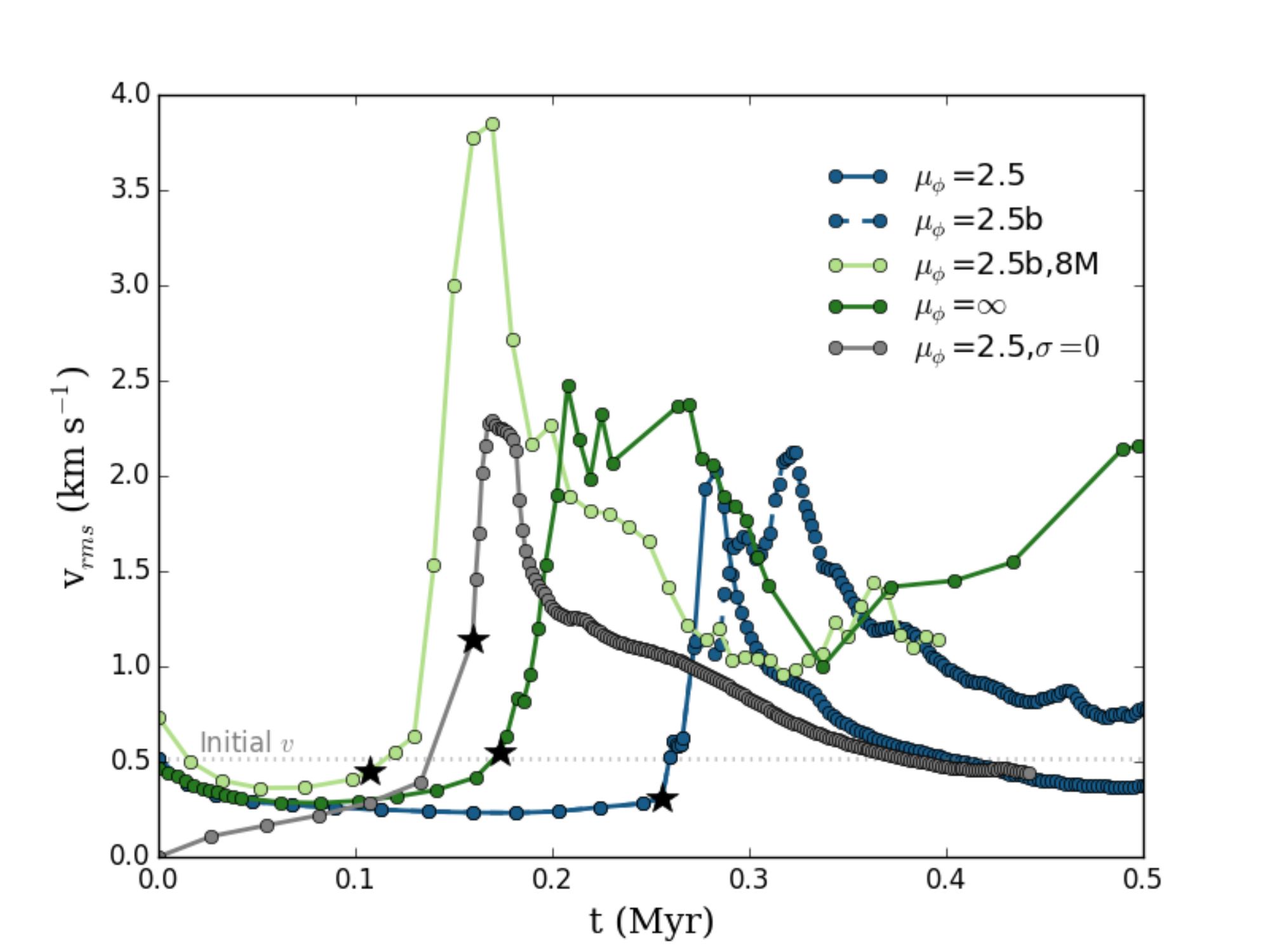}
\end{center}
\caption{ Mass-weighted 3D gas velocity dispersion as a function of time. Stars mark the time of protostar formation.  The dashed gray line denotes the initial velocity dispersion.  Top: velocity dispersion for various mass-to-flux ratios. Bottom: velocity dispersion comparison for two different initial masses, with and without binarity (denoted by a ``b"), and cases with no magnetic field or turbulence.
\label{vvst} }
\end{figure}

\begin{figure}
\begin{center}
\vspace{-0.2in}
\includegraphics[scale=0.46]{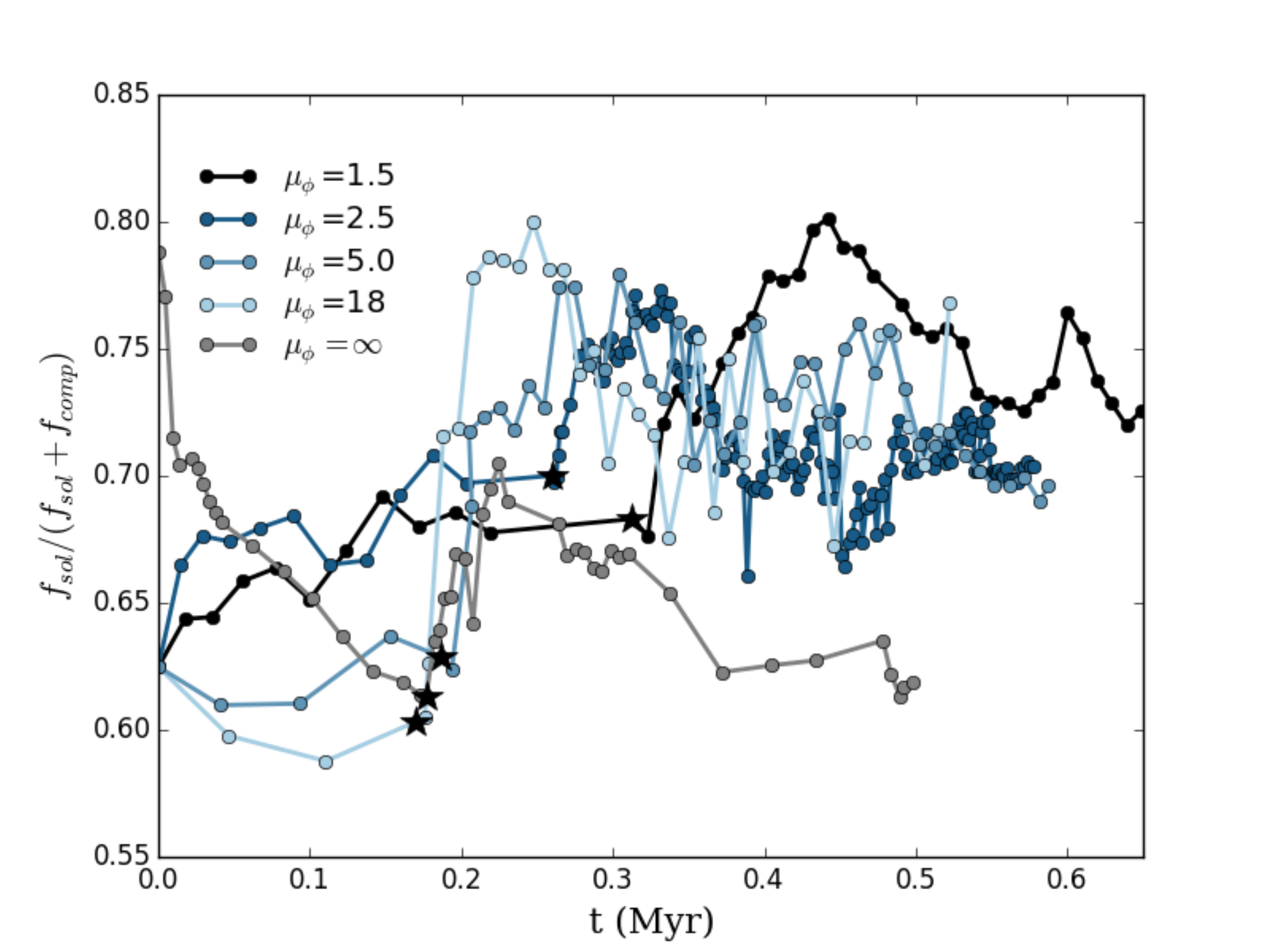}
\end{center}
\caption{The fraction of solenoidal velocity motion in the dense gas ($n_{\rm H}\geq 10^4$ cm\eee) for various mass-to-flux ratios. Stars mark the time of protostar formation.
\label{solvst} }
\end{figure}

\begin{figure*}
\begin{center}
\hspace{-0.1in}
\includegraphics[scale=0.20]{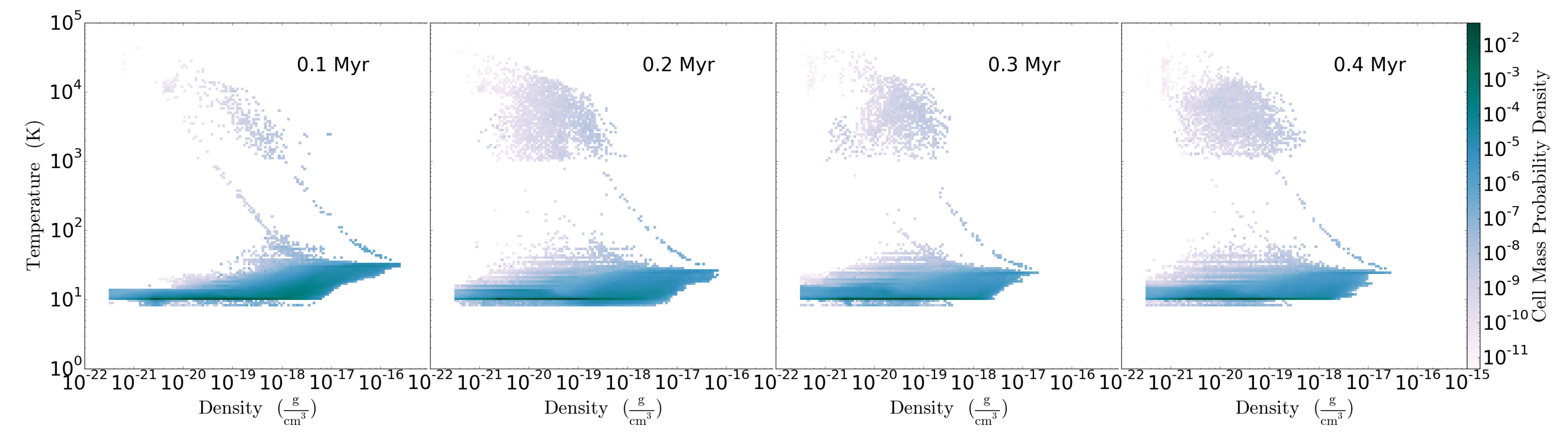}
\end{center}
\vspace{-0.1in}
\caption{ Gas temperature as a function of gas density for run M4P2.5, where the time as measured from the formation time of the protostar is noted in each panel. The colorscale indicates the mass fraction at a given temperature and density.
\label{temp} }
\end{figure*}

\subsubsection{Turbulence}\label{turb}

In all cases, the outflow has a strong impact on the core energetics and it drives a significant amount of turbulence, loosely defined as disordered motions in the dense gas. Figure \ref{vvst} shows the rms velocity dispersion as a function of time.  As the cores collapse, the initial turbulent motions gradually decay. The dispersion increases slightly just before protostar formation as a result of gravitational infall. Once the outflow begins the global dispersion increases sharply. This $\sim 0.05$ Myr period corresponds to both the highest protostellar accretion (see \S\ref{acc}) and the time during which the outflow breaks out of the core. The dispersion remains high for $\sim 0.1$ Myr, a timescale similar to the observed Class 0 lifetime \citep{evans09}.  Once most of the core mass is accreted or expelled the velocity dispersion declines by a factor of 2-4 and the velocity dispersion approaches $\sim$0.5-1 $\kms$. Inspection of the images in Figure \ref{vr_Density} shows that the velocity dispersion is not wholly dominated by the outflow. Instead, the interaction between the outflow and core drives significant secondary motion in gas that is not directly part of the jet or molecular outflow. We quantify this effect in \S\ref{outdriving}. Thus, as a result of outflow activity turbulence is maintained after protostar formation and lasts for several 0.1 Myr.

The velocity evolution including magnetic fields is similar to that of the unmagnetized calculations of OA14. The bottom panel of Figure \ref{vvst}  shows that the velocity dispersion tends to be higher and remain so when $B=0$.  However, due to magnetic field amplification, the evolution of the initially weak field case of $\mu_\phi=18$ at late times is more similar to the stronger field cases. This may suggest that magnetic fields damp some of the turbulence generated by the outflow. The higher mass core (run M8P2.5b), which forms a higher mass protostar with a stronger outflow, achieves and sustains a higher global velocity dispersion.  Run M4P2.5b, which follows the evolution of the close binary, achieves a velocity dispersion twice that of the run in which the binary components are merged. This suggests that multiple outflows, especially when misaligned \citep[e.g.,][]{Offner16}, can efficiently regenerate local turbulence. The 8$\msun$ core also forms a binary, which may also contribute to its higher velocity dispersion. 

The details of the outflow properties may impact the efficacy of energy injection, so these results depend somewhat on the outflow model. For example, OA14 explored the effect of changing the outflow angle on the core evolution. However, they found only a modest difference in the velocity dispersion between models with $f_w=0.2,0.3$ and $\theta_0=0.01-0.1$. Wider angles than explored in this work may produce a qualitatively different result.

A velocity field can be decomposed into two orthogonal components: a compressive field ($\vec \nabla \times \vec v_c = 0$), which reflects the amount of divergence in the velocity field, and a solenoidal field ($\vec \nabla \cdot \vec v_s = 0$), which represents the amount of circular motion or ``stirring."  To calculate these two components, we flatten the velocity field to a uniform $256^3$ grid, which encompasses data up to two levels of AMR refinement. We investigate the relative motions in the denser gas by masking out material with densities $n_{\rm H_2}< 10^4$ cm$^{-3}$.

Figure \ref{solvst} shows the fraction of solenoidal motion in the dense gas as a function of time. Gravitational collapse and outflow interaction with the core affect the relative strength of the two components.  Gravity and turbulence together produce a solenoidal fraction of $\sim 0.65$ or about twice as much solenoidal as compressive motion.  Once the outflow launches, the fraction of solenoidal motion increases to $\sim 0.75$ or 3:1. This increase is consistent with prior analysis of outflow-driven turbulence in clusters \citep[OA14,][]{hansen12}. The ratio declines again after the main accretion and outflow phase.
We find the solenoidal fraction for all gas, including lower density material, is similar to that of the dense gas (usually within 20\%). 

 Gravitational collapse may also drive turbulence \citep{robertson12,murray15}. This would probably manifest as a change in the solenoidal fraction around the time of protostar formation. 
Figure \ref{solvst} shows the solenoidal fraction remains fairly constant in the MHD runs during the starless phase, although there is a slight increase during $\sim 0.1-0.2$~Myr.  Run M4P$\infty$ shows a monotonic decline in solenoidal fraction, which suggests that changes to the magnetic field rather than gravity specifically more effectively drives turbulence.

\citet{federrath12} demonstrated that the star formation rate in simulated molecular clouds depends strongly on the nature of the turbulent driving, with more compressive turbulence producing more efficient star formation. This is supported by analysis of local molecular clouds  by \citet{orkisz17}, who find regions with more solenoidal motion have lower star formation rates. \citet{brunt14} show that the fractional power in the solenoidal field converges to $0.7\pm 0.1$ for high-mach number turbulence independent of driving. This value is consistent with our finding in the prestellar core phase, although our turbulence is transsonic and the driving force is gravity. The higher solenoidal fraction in our protostellar cores underscores that outflows not only entrain and remove dense gas but also stir their surroundings, which inhibits rather than promotes star formation within the core.

The energy and velocity power spectra provide an alternative means to assess the impact of outflows on turbulence. Simulations show outflows can produce differences in slope or features in the power spectrum \citep[e.g.,][]{carroll09,carroll10}. Stronger energetic feedback from winds, which evacuate molecular material on parsec scales, may produce even more significant changes \citep{Offner15,boyden16}. However, detecting and characterizing the influence of feedback in observational spectral data is challenging and such studies have focused on larger scales than those studied in this work \citep[e.g.,][]{swift08}. 


\begin{figure}
\begin{center}
\includegraphics[scale=0.46]{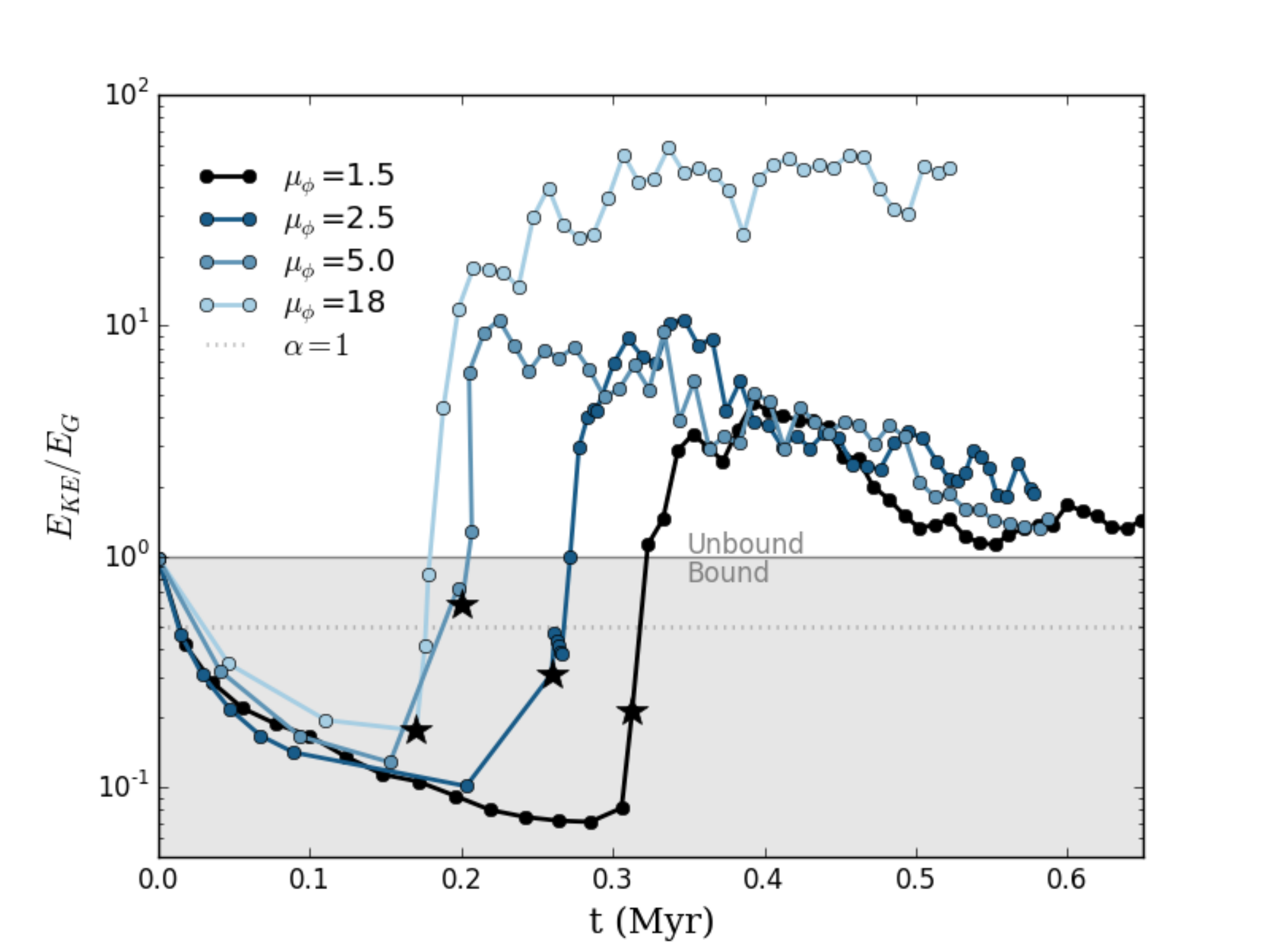}
\end{center}
\caption{The ratio of kinetic to gravitational energy of the dense gas ($n_{\rm H}\geq 10^4$ cm\eee) as a function of time for various mass-to-flux ratios. Stars mark the time of protostar formation. The gray shaded region indicates the ratios for which the gas is gravitationally bound. The dotted line shows the virial parameter $\alpha_{\rm vir} = 1$ for a uniform sphere.
\label{boundvst} }
\end{figure}

\begin{figure*}
\begin{center}
\includegraphics[scale=0.46]{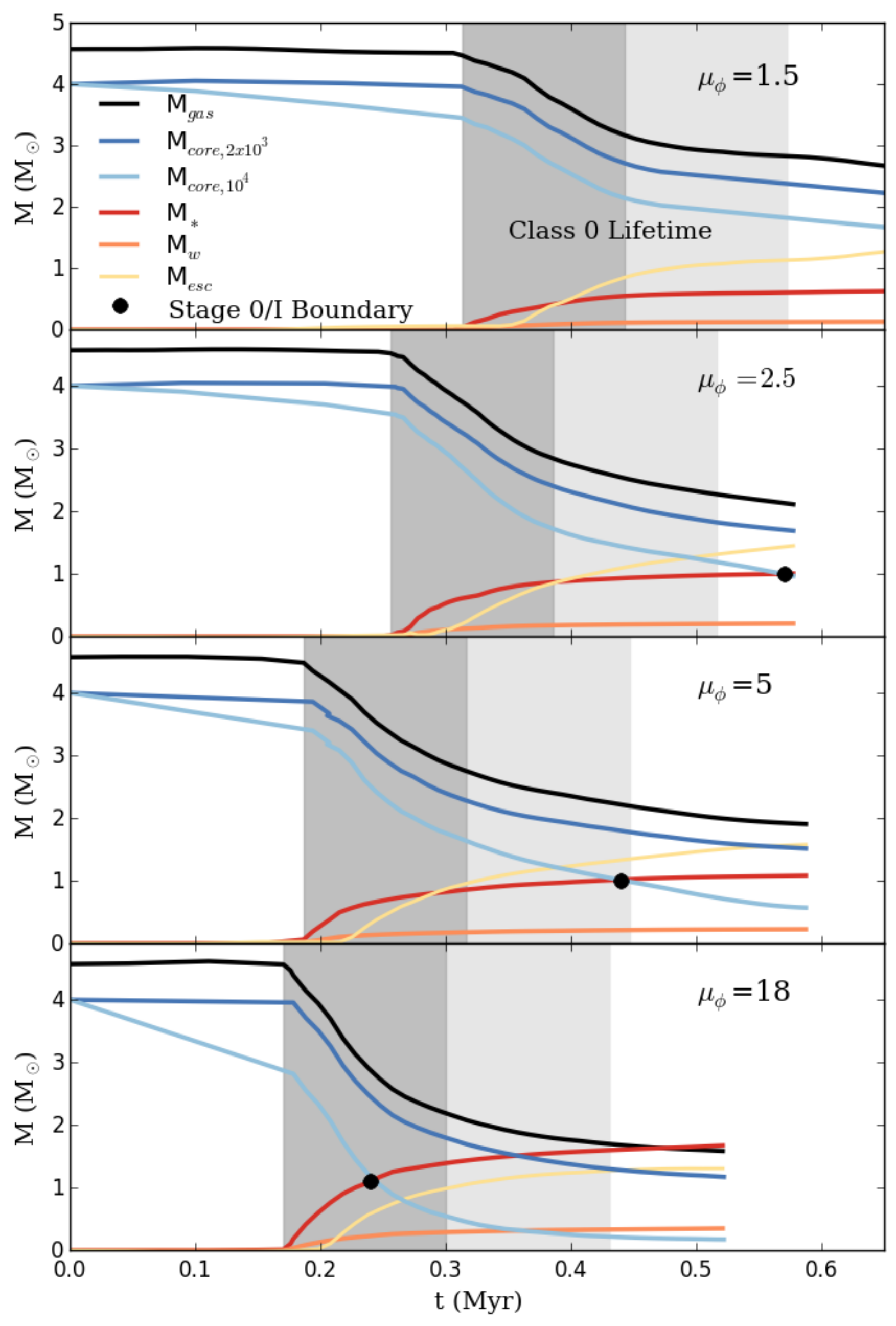}
\includegraphics[scale=0.46]{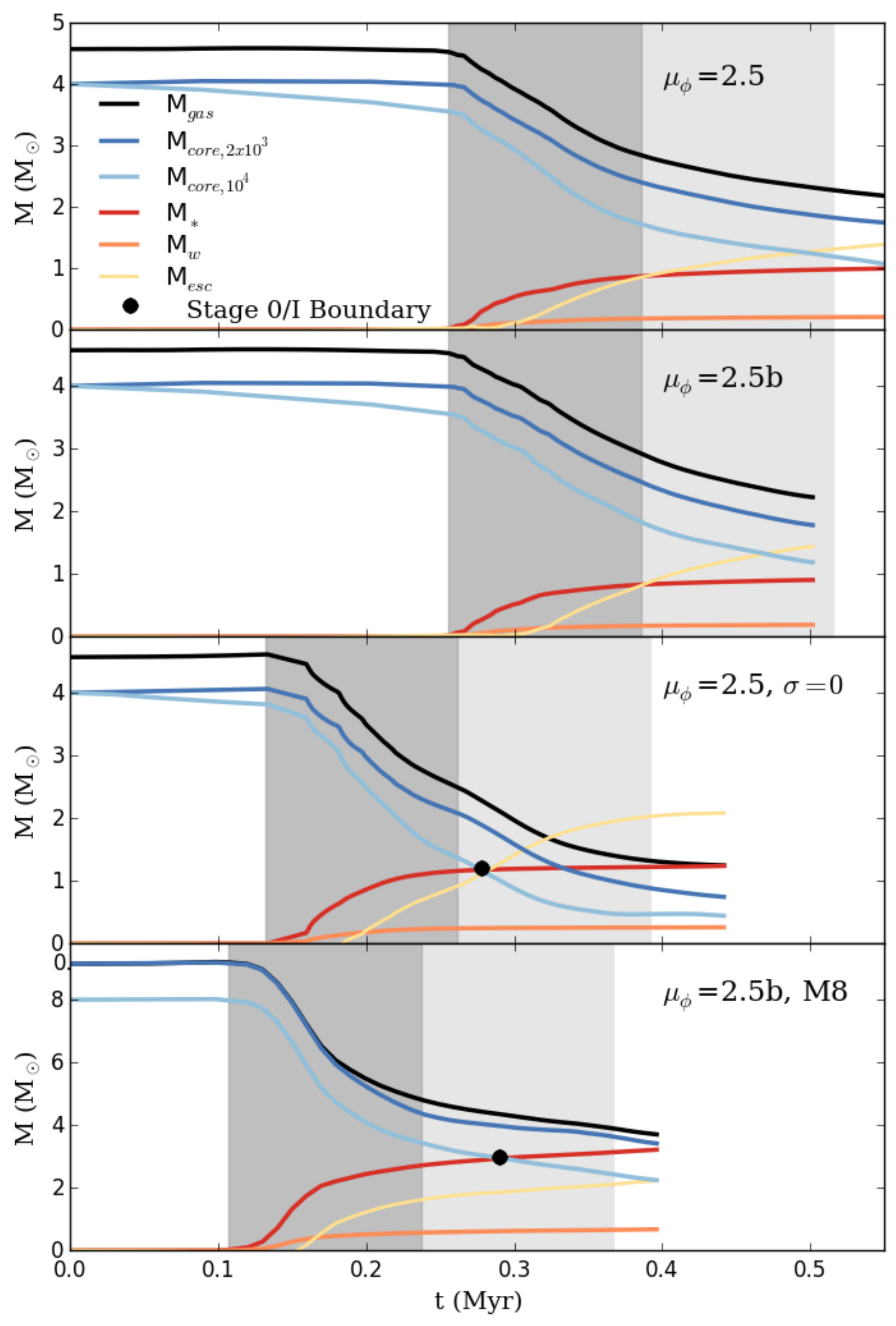}
\end{center}
\caption{Time evolution of the total gas mass ($M_{\rm gas}$), gas mass with $n_{\rm H}\geq 2 \times 10^3$ cm\eee ($M_{\rm core, 2\times 10^3}$), gas mass with $n_{\rm H}\geq 10^4$ cm\eee ($M_{\rm core,10^4}$), total mass launched in the jet ($M_w$), protostellar mass ($M_*$), and mass that has escaped from the domain ($M_{\rm esc}$). The gray shaded region indicates the observed Class 0 lifetime ($t_0=0.13-0.26$ Myr) marked from the time of protostar formation  \citep{dunham15}. The black dot indicates the point at which the protostellar mass equals the dense gas mass, which is the Stage 0/I boundary by definition.
\label{mvst} }
\end{figure*}

\subsubsection{Gas Temperature}\label{sec_temp}

 {\sc orion2} follows the evolution of the gas temperature by solving the radiation diffusion equation. Dust is the predominant coolant, but atomic cooling does occur in the dust-free hot outflow material.  Figure \ref{temp} shows the evolution of the gas temperature for run M4P2.5. Gas is warm because it is heated locally by the forming protostar, it is associated with the hot outflow or it is gas entrained by the outflow. Gas heated by the protostar lies predominately at higher densities and exhibits temperatures of $\sim$20-50~K (locus of points in the bottom right of each panel in Fig.~\ref{temp}). Since the densest gas 
lies within the sphere of influence of the protostar, there is no high-density $\sim$10~K gas. As demonstrated by \citet{Offner09}, the inclusion of radiative feedback in calculations of low-mass star formation significantly enhances heating above that expected for a typical equations of state \citep[e.g.,][]{masunaga00}. By construction, the outflow is assumed to be ionized, and outflowing gas is deposited on the domain with a temperature of $10^4$~K; this gas lies in the top center of the panels in Fig.~\ref{temp}. As the outflow mixes with the envelope the temperature declines, and the bulk of the entrained material inhabits intermediate temperatures and densities ($\rho \sim 10^{-20}-10^{-17}$~g~cm$^{-3}$). 

The amount of hot gas, $T\sim 10^3-10^4$~K, increases as the outflow cavity widens. At later times, turbulent mixing between the outflow and envelope increases the distribution of intermediate and low-density warm gas. The presence of warm, high-density material remains similar; at early times the high accretion luminosity drives the heating, while at later times the intrinsic luminosity of the protostar dominates. The other models follow a similar progression.

\subsubsection{Boundedness}\label{bound}

Cores are considered gravitationally bound when their gravitational energy exceeds their total kinetic energy. We define the fraction of kinetic to gravitational energy as
\beq
\frac{E_{\rm KE}}{E_{\rm G}} = \frac{\frac{1}{2}M_{\rm core} V_{\rm core}^2}
{\int \rho r \left( \frac{d\Phi_{\rm G}}{dr} \right) \hat e_r dr^3 },
\label{ratio}
\eeq
where $V_{\rm core} = \sqrt{3}\sigma$ is the 3D velocity dispersion, $\phi_G$ is the gravitational potential calculated directly by {\sc orion2} and the integral is computed as a sum over concentric shells.
 This simplification excludes the contribution of the magnetic field, which also contributes to stability. We restrict the analysis to the kinetic and gravitational energies, since it is an approximation used pervasively in the star-formation literature. 

For a sphere with uniform density, the gravitational energy is $E_{\rm G} = 3 G M^2/(5R)$. Eqn.~\ref{ratio} can be related to the virial parameter, $\alpha_{\rm vir}$:
\beq
\frac{E_{KE}}{E_G} = \frac{5 \sigma^2 R}{2GM} = \frac{1}{2}\alpha_{\rm vir}.
\eeq
Thus, all cores begin with $E_{\rm KE}/E_{\rm G}=1$.

Figure \ref{boundvst} displays the ratio of kinetic to gravitational energy in the dense gas as a function of time. In all cases, boundedness increases during the initial collapse, but the core as a whole quickly becomes unbound once the outflow begins. There is good correspondence between the magnetic field strength and boundedness, with more magnetized cores remaining more bound at later times. 

Note that the calculation of the gravitational energy excludes the protostellar mass, so $E_{\rm KE}/E_{\rm G}$ represents an upper limit. These virial parameters are similar to those for observed cores, which can exceed $\alpha \simeq 10$  \citep{kauffmann13}. However, we note the outflow contributes significantly to the dispersions in the weaker magnetic field cases (see \S\ref{outdriving}).  Our estimate also excludes the effect of external pressure, which some studies conclude has a large impact on core boundedness \citep[e.g.,][Kirk et al. 2017]{bertoldi92,field11,pattle15}. 

\begin{deluxetable}{lcccc}
\tablecolumns{10}
\renewcommand{\tabcolsep}{0.07cm}
\tablecaption{Model Results \label{starprop}}
\tablehead{ \colhead{Model\tablenotemark{a}} &  
  \colhead{$M_{\rm *,f}$($\msun$) } &
   \colhead{$\epsilon_{\rm max}$} & 
 \colhead{$M_w$($\msun$) } &
 \colhead{$\dot M_w v$($\msun$ kms$^{-1}$) } } 
\startdata
M4P1.5     & 0.64 & 0.16 & 0.13  & 5.1    \\ 
M4P2.5     & 1.01 & 0.25 & 0.21  & 8.3   \\ 
M4P2.5b   & 0.91 & 0.23 & 0.19   & 6.8    \\ 
M4P2.5nt  & 1.24 & 0.31 & 0.26   & 10.2  \\ %
M4P5      & 1.08 & 0.27 & 0.23   & 8.0   \\ 
M4P18     & 1.68 & 0.42 & 0.35   & 12.7   \\ 
M4P$\infty$ & 1.61 & 0.40 & 0.34   & 15.8   \\ 
M8P2.5b   & 2.94 & 0.37 & 0.62  & 22.44    
\enddata
\tablenotetext{a}{The model name, final protostellar mass, final efficiency, total mass launched by the outflow, and total momentum launched by the outflow. The efficiency is defined as $\epsilon=M_{\rm *,f}/M_{\rm core}$. This is a lower limit since some residual gas remains on the domain. For M4P2.5b and M8P2.5b totals include the contribution from the secondary.}
\end{deluxetable}

\subsection{Protostellar Properties and Evolution}\label{protoprop}

\subsubsection{Envelope Processing and Dispersal}\label{envel}

Gas in the core has four possible evolutionary trajectories. It can accrete onto the protostar, $M_*$; it may be ejected in a high-velocity jet as described by the outflow sub-grid model, $M_w$; it can be entrained by the jet and ejected from the simulation domain, $M_{\rm esc}$; or it may remain  as molecular material, $M_{\rm core}$. Any gas remaining at the end of the calculation will eventually follow one of the other three evolutionary paths. Figure \ref{mvst} shows each of these quantities as a function of time. The amount of remaining dense gas is defined using thresholds of $n_{\rm H} \geq 2 \times 10^3$ cm$^{-3}$ ($M_{\rm core,2\times 10^3}$) and $n_{\rm H} \geq 10^4$ cm$^{-3}$ ($M_{\rm core,10^4}$).

The protostar masses formed within the $4\msun$ cores at the end of the calculation span $\sim$0.6-1.7 $\msun$ (see Table \ref{starprop}), with stronger field calculations producing proportionally smaller protostars. This occurs because the magnetic field introduces asymmetry into the accretion process. Gas continues to freely accrete along the field lines, where there is no magnetic resistance. Along this direction, accretion proceeds similarly to gas in non-magnetized models. In contrast, gas may only accrete in the direction perpendicular to the field by bending the magnetic field lines, which is subject to a magnetic tension force. During the first 0.1~Myr after protostar formation, self-gravity is able to overcome the magnetic restoring force and the field is dragged inward, creating a clear hour-glass shape in the field lines \citep[see Fig.~2 in][]{lee17}. This signature appears in observed magnetic polarization measurements of young protostellar sources \citep[e.g.,][]{girart06,stephens13,hull14}.
However, accretion remains strongly inhibited compared to the non-magnetized case. The net result is that accretion occurs more slowly in the more magnetized simulations, and significant core gas remains on the domain even 0.3~Myr after the protostar forms.

Gas entrainment and expulsion are also affected by the magnetic asymmetry. Smaller protostars launch weaker outflows, which entrain less material. However, the gas in the magnetized calculations is less centrally concentrated, which makes the envelope material easier to unbind. These effects appear to cancel, and 0.3~Myr after protostar formation, all four calculations have ejected $\sim 1.5 M_\odot$ from the domain, and thus, all have similar escape fractions. 

The core mass defined using the higher density threshold, $M_{\rm core,10^4}$, declines before the protostar forms as a result of turbulent mixing with the ambient medium.  The initial turbulence has the largest impact on the amount of dense gas when the field is weakest.
The core mass defined using the lower density threshold, $M_{\rm core,2\times 10^3}$ remains around $4 \msun$ and only declines at protostar formation, once gas begins to be accreted or expelled. This suggests that the lower threshold tracks and distinguishes the initial core gas from the ambient medium gas.  The amount of residual core gas for both thresholds reflects the slower accretion in the more magnetized runs.

As shown in the second set of panels in Figure \ref{mvst}, the more massive core and the core without initial turbulence collapse more quickly. These protostars grow more rapidly and reach higher masses despite having the same initial mass-to-flux ratio. This underscores the utility of an effective mass-to-flux ratio that takes into account the thermal pressure, as proposed by \citep{mckee89}. These results also show that the collapse and subsequent accretion depends on the amount of turbulence in the core.


\subsubsection{Stage 0 Lifetime}\label{stage0}

The Stage 0 lifetime, which encompasses the main accretion phase, is defined to span from protostar formation until the protostellar mass exceeds the envelope mass, $M_{\rm core}>M_{*}$. Theoretical models and simulations have estimated Stage 0 lifetimes ranging from $\sim$0.1-0.3 Myr \citep{young05,machida13}.  
However, these estimates depend on the details of the initial conditions and are based on models that have neglected one or more of turbulence, magnetic fields and outflows.

Determining evolutionary stage directly from observations is challenging.
Given the heavy obscuration of young protostars, only indirect indicators are available to estimate the protostellar age and mass.
For example, the shape of the spectral energy distribution (SED) of protostellar cores provides an estimate of evolutionary stage \citep{dunhamppvi14}. During the earliest period of evolution, radiation from the embedded protostar is heavily reprocessed by dust and the SED peaks in the far infrared. This ``Class 0" stage serves as an observational analog to the Stage 0 phase.  Studies of local star forming regions find Class 0 lifetimes of $t_0=$0.13-0.26 Myr \citep{dunham15}. Stage 0 and Class 0 are sometimes used interchangeably; however, exactly how a given SED correlates with underlying protostellar properties, including age, mass and accretion rate is uncertain. SEDs are strongly influenced by geometry, viewing angle and accretion variability \citep{robitaille07,Offner12,dunham12}. 
The Class 0 lifetime also depends on an assumed disk lifetime, which is used to calibrate the length of the Class 0, I and II stages \citep[e.g.,][]{evans09}. 

Figure \ref{mvst} compares the observed Class 0 lifetime with the Stage 0 lifetime for our simulation models.
The black circle indicates the transition point between Stage 0 and Stage I for each model, while the gray shading indicates the range of Class 0 lifetimes \citep{dunham15}. We adopt the higher density threshold to define the Stage 0 lifetime, since it is more consistent with the densities of observed cores \citep{andre00}.  For the two weaker field models, the Stage 0 lifetime is within the observed Class 0 lifetime. In the weakest field case, the Stage 0 lifetime, $t_0\sim$0.15~Myr, is similar to that found by OA14 for non-magnetized runs. In the strongest field model, the core mass and envelope mass 
do not reach parity before the end of the calculation, which implies a Stage 0 lifetime exceeding $0.35$ Myr. 

The Stage 0 lifetime for the 8$\msun$ core is $\sim 0.18$Myr, which is half that of the 4$\msun$ with the same initial mass-to-flux ratio. This reflects the lesser influence of thermal pressure in the more massive core, which consequently has a higher effective mass-to-flux ratio. Model M4P2.5b, which contains a binary, has a very similar evolution to model M4P2.5, so that it appears its Stage 0 lifetime will be similar even though the calculation does not reach the Stage 0/I transition. A more detailed statistical study over a range of binary properties is necessary to draw general conclusions about how multiplicity impacts protostellar evolution and the Stage 0 lifetime.

Past work including more detailed SED modeling indicate the Stage 0 and Class 0 lifetimes are comparable on average \citep{robitaille07,Offner12}, although these models neglected magnetic fields. If accretion is variable, then the Stage and Class lifetimes may be poorly correlated \citep{dunham12}. Since the Class 0 lifetime is also calibrated using the assumed disk lifetime, this introduces another degree of uncertainty. Recent pre-main sequence models suggest Class II objects may be $\sim$50\% older than previously thought \citep{bell13}. Our results may imply that the Stage 0 lifetime is longer than the Class 0 lifetime because the relationship between Stages and Classes is poor or that fields in dense cores are weak ($\mu_\Phi >2.5$). The former possibility would mitigate the ``luminosity problem'', in which protostars are observed to be dimmer than expected given assumptions for the average accretion rate \citep[see \S\ref{lum}]{dunhamppvi14}. 
Note that ongoing accretion from the environment, which is not included in these calculations, would prolong the Stage 0 lifetime by increasing the dense gas reservoir. Other physical differences not explored in these models, like initial density structure and rotation, could also impact the Stage 0 lifetime. 

\begin{figure}[t]
\begin{center}
\vspace{0.1in}
\includegraphics[scale=0.48]{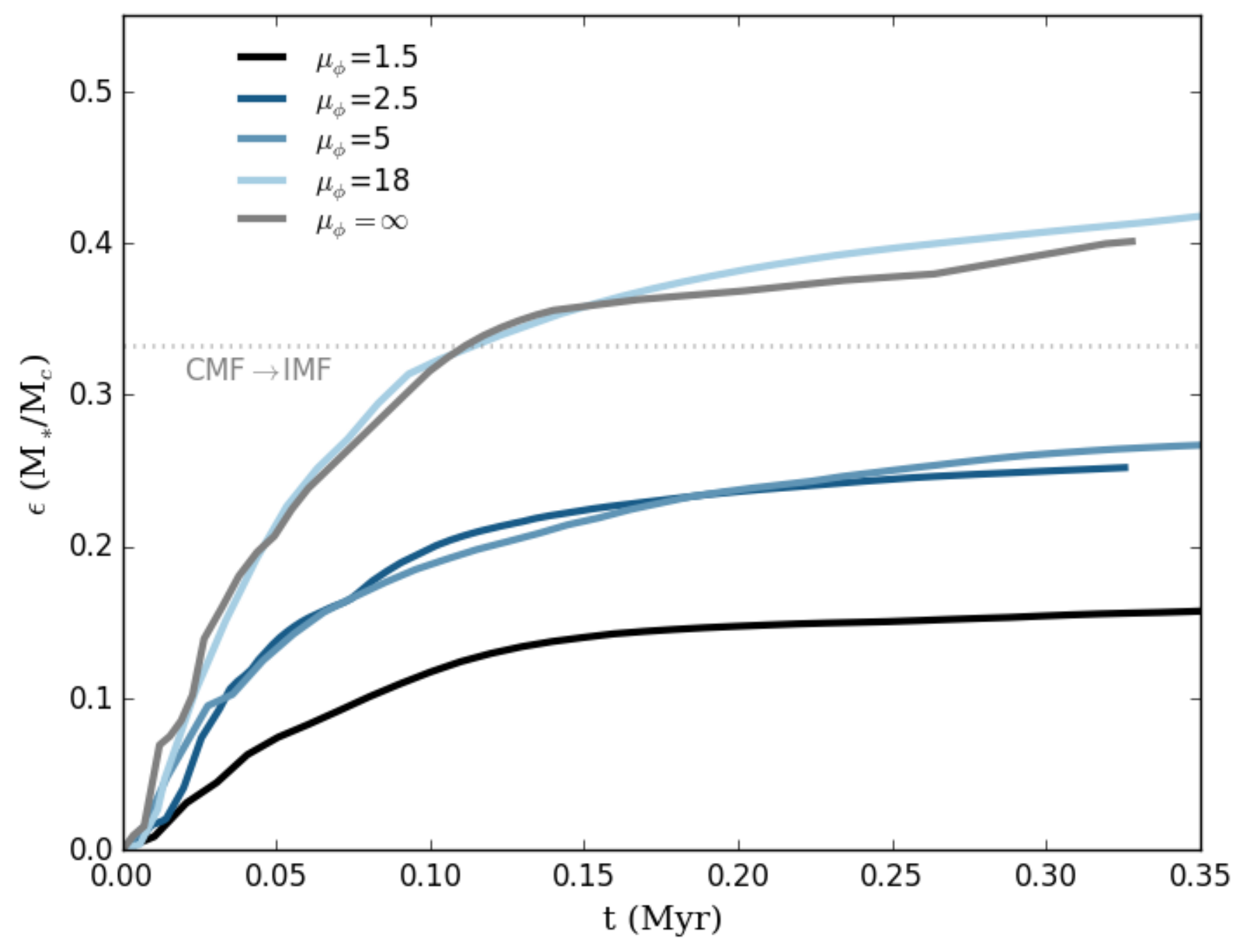}
\includegraphics[scale=0.48]{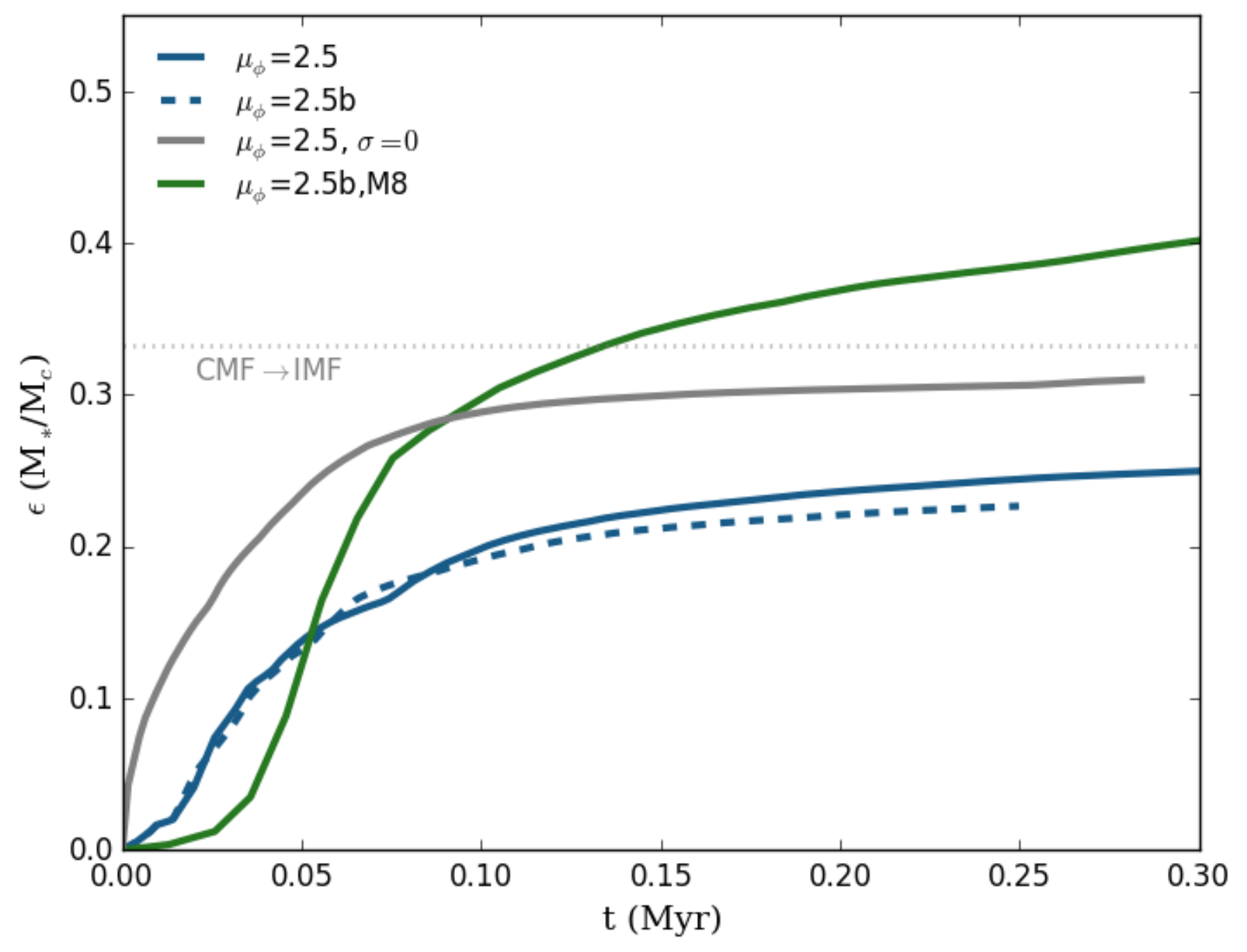}
\end{center}
\caption{ Top: Efficiency as a function of time for different mass-to-flux ratios, $\mu_\phi$. Bottom: Efficiency as a function of time for two different core masses with the same mass-to-flux ratio. Dashed line indicates the runs with binaries. The dotted line indicates $\epsilon=0.33$, the apparent offset between the stellar initial mass function (IMF) and the dense core mass function (CMF).
\label{effvst} }
\end{figure}

\subsubsection{Star-Formation Efficiency}\label{eff}

Outflow entrainment together with magnetic pressure support prescribe the star formation efficiency of the core. We define the efficiency, $\epsilon$, as the ratio of the protostellar mass to the initial dense core mass. The top panel in Figure \ref{effvst} shows the efficiency as a function of time for different field strengths. 
The figure illustrates the strong relationship between the initial magnetic field and the star formation efficiency: as the field increases, the efficiency declines by more than a factor of two (see also Table \ref{starprop}). 
This occurs because larger initial fields more significantly inhibit accretion perpendicular to the field lines. As a result, the protostar grows more slowly at all times. 

Altogether, the final efficiencies range from $\sim 15$\%-40\%. This is slightly lower than the 25\%-75\% estimated by \citet{matzner00} for individual cores. Final stellar masses may increase slightly with additional numerical evolution, however the slopes of the curves in Figure \ref{effvst} suggest that most runs have converged and little additional accretion will occur.

We find the efficiency of the weakest field case, $\mu_\phi=18$, is similar to that of the non-magnetized simulation. As shown in Figure \ref{vr_Density}, the gas remains fairly spherical and, consequently,  collapse is not strongly impeded by the magnetic field. 
Although the field delays collapse in the $\mu_\phi=2.5$ run (e.g., Fig. \ref{mvst}), thereafter the evolution proceeds similarly to the $\mu_\phi=5$ case and the final efficiencies are similar. This suggests that the efficiency depends on the evolution of the rms field and effective Alf\'ven Mach number rather than the exact value of the initial field.

The bottom panel in Figure \ref{effvst} compares runs with initial core masses of 4\msun ~and 8\msun, which have the same mass-to-flux ratio. The higher mass core concludes with a higher efficiency; however, it has a higher effective mass-to-flux ratio, $\mu_{\phi,\rm eff}$,  and also contains a multiple system. 
The second protostar forms around 0.05 Myr,
and the interaction between the two, which migrate from separations of $\sim 3,000$~AU to 100~AU, may enhance accretion \citep[c.f. Fig.~1 in][]{Offner16}. However, the multiple system in the $4\msun$ case, which also migrates, appears  
to have little impact on the final efficiency. 

The dotted horizontal line in Figure \ref{effvst} indicates the offset between the stellar initial mass function (IMF) and the observed core mass function \citep{alves07,enoch08,konyves15}. The significance of the CMF and IMF shapes and their apparent constant offset remain debated \citep{Offner14ppvi}: the similarity could either indicate that stellar masses are inherited from their parent cores or that the resemblance is a coincidence caused by random processes. The correspondence between the simulated efficiencies and the value inferred from the CMF-to-IMF mapping at minimum suggests that outflow entrainment is an important factor in the final star formation efficiency. Since these simulations represent isolated cores and do not consider any additional flow of material onto the core \citep[e.g.,][]{smith09,kirk13}, the protostellar masses likely represent lower limits.

 The final efficiency may be sensitive to the details of our model for protostellar outflow launching. 
For example, the outflow direction changes by tens of degrees over the course of the main accretion period \citep{lee17}, which may enhance entrainment. An alternative model could set the outflow launching direction parallel to the normal of the accretion disk rather than the normal of the protostellar spin. \citet{fielding15} considered this possibility and found that the change in disk orientation over time was larger than that set by the stellar spin. 
 Recent ALMA studies have investigated the opening angle and velocity distribution on $\sim 10$ au scales: Class 0 source HH46/47 has both jet-like and wide-angle components \citep{zhang16}, while the outflow of Class I source TMC1a is best described as a wide-angle disk-wind \citep{bjerkeli16}. Thus, the launching mechanism may also vary among sources and over time.

\subsubsection{Multiplicity}\label{multiplicity}

The occurrence of gravitational fragmentation in the simulations depends on core mass, magnetic strength and initial turbulent seed. In runs M4P2.5b and M8P2.5b binary formation arises from the fragmentation of small-scale filaments created by the initial core turbulence \citep{dunham16,Offner16} rather than fragmentation within a gravitationally unstable disk  \citep{kratter10,tobin16b,lewis17}. 

The dominant impact of magnetic fields on fragmentation remains debated. Strong magnetic fields may promote fragmentation by producing a flattened dense structure with sub-structure prone to fragmentation \citep{fischera12,pineda15}. Cores with weaker fields are more symmetric and centrally condensed, which tends to reduce the incidence of protostellar multiplicity, an idea first proposed by \citet{boss00}.  Fragmentation also depends on the turbulent seed and core mass as discussed in \citet{Offner16}, who found massive, turbulent cores are more prone to fragmentation. In contrast, \citet{myers13} found that magnetic fields suppressed fragmentation on $\gtrsim 1000$ AU scales in simulations of very massive star-forming cores.


The binary separations of both M4P2.5b and M8P2.5b were analyzed in \citealt[][(M4 and M8 in their Fig.~1)]{Offner16}.
The binaries have initially wide separations with $d \gtrsim 500$ AU.  However, the pairs quickly migrate to closer separations.  

In both cases, the binary mass ratio at the end of the calculation is $\sim 9$. The pairs maintain a similar mass ratio throughout their evolution, preserving their early mass differences.  In non-magnetized calculations, the later-forming secondary can gain mass quickly if its orbit is within the accretion disk of the primary \citep[e.g.,][]{kratter10}. Since our calculations do not resolve accretion disks, the secondaries cannot accrete mass via this channel. Accretion disks are also expected to be more compact when magnetic affects are included \citep{lippvi14}, so binaries with separations $>50$ AU may not have a shared accretion disk. 

 \citet{tobin16b} found that Class 0 sources in Perseus have a multiplicity fraction (MF) of $0.57 \pm 0.09$, where the MF is defined as the ratio of the number of multiple systems to the total number of systems. \citet{Offner16}, who modeled 12 cores with masses from 4-8 \msun, different turbulent seeds and initial mass-to-flux ratio of 2.5, found $ MF = 7/12 = 0.58 \pm 0.15$, which is in excellent agreement with the observations of young protostars. Fewer multiples are produced in this study (MF $= 2/7 \simeq 0.3$), but given that our core properties are not intended to statistically sample the properties of observed cores and the small number of statistics in both studies, these MF fractions should not be over-interpreted. To date, few numerical studies have investigated binary formation for solar-type stars including magnetic fields \citep{machida08,myers14,Offner16,wurster17,li17} and, consequently, the statistics are poor. Future work is needed to address this.

\begin{figure}
\begin{center}
\includegraphics[scale=0.45]{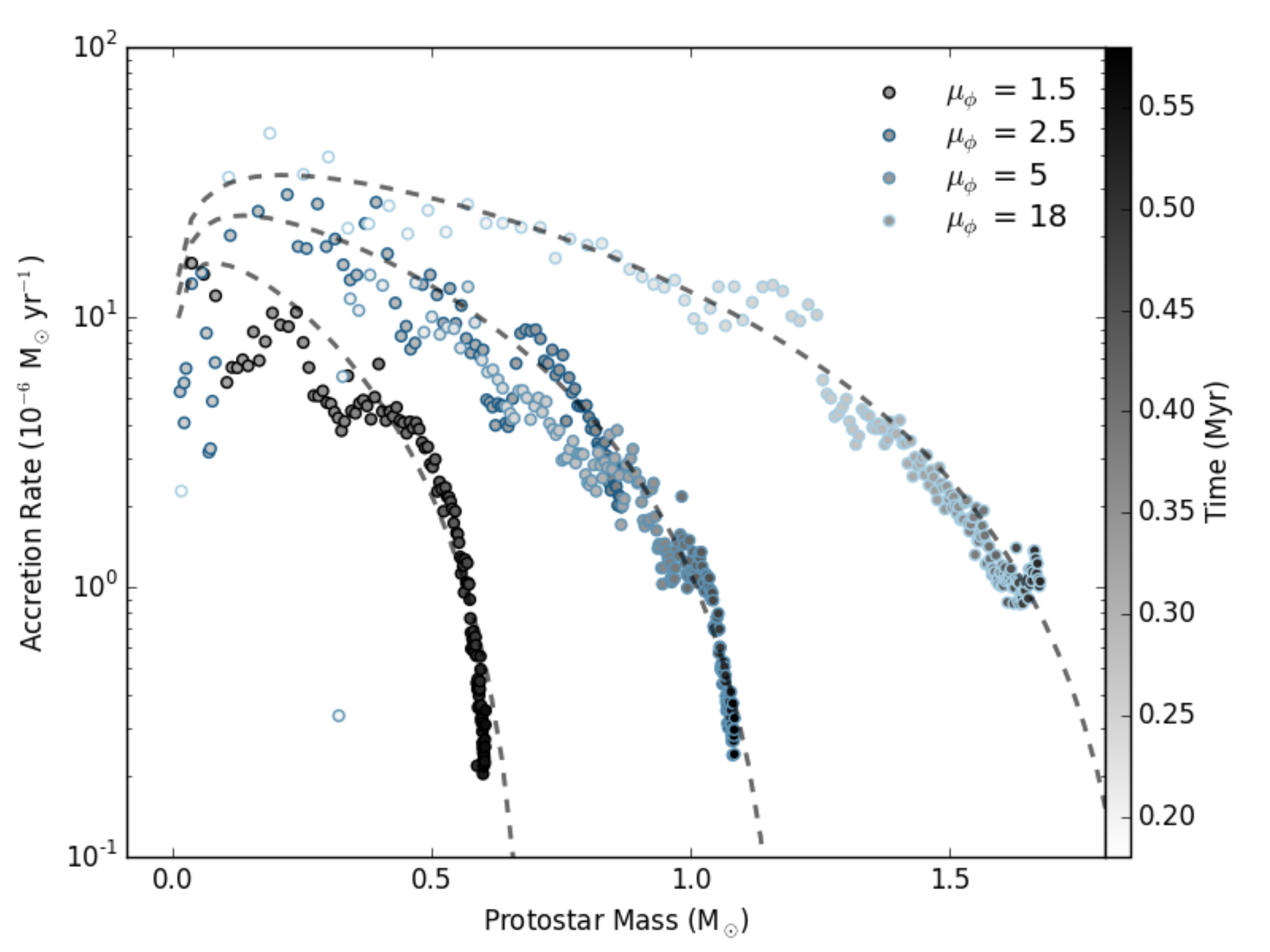}
\end{center}
\caption{ Averaged protostellar accretion rate as a function of protostellar mass. The colorbar indicates the time. The dashed lines are the predicted accretion rate according to Eqn.~ \ref{tcfit}, where $\dot m_{\rm TC} = 1.4 \times 10^{-4} M_\odot$  yr$^{-1}$ and $m_f = 0.7, 1.2, 1.9~ M_\odot$ from lowest to highest $\mu_\phi$. Runs M4P2.5 and M4P5 have approximately the same final mass and compare well with the $m_f=1.2$ model. \label{mdotvsmp} }
\end{figure}

\subsubsection{Protostellar Accretion}\label{acc}

Figure \ref{mdotvsmp} shows the protostellar accretion rate as a function of protostellar mass for the four magnetized models. All the models have high accretion rates for the first $\sim$0.1 Myr with peak values inversely proportional to the initial magnetic field strength. Thereafter, accretion declines sharply for the stronger field case and more gradually for the weaker field cases. This is due to the magnetic field inhibiting accretion perpendicular to the field lines. Runs M4P2.5 and M4P5 have similar accretion histories since their Alf\'ven Mach numbers are comparable during the first $\sim$0.15 Myr of the protostellar lifetime (see Fig.~\ref{mavst}). 

For comparison we over-plot the accretion rate given by the tapered turbulent core (TC) model on Figure \ref{mdotvsmp} \citep{mckee03,mckee10}. The TC model predicts the accretion rate is a strong function of the instantaneous mass, $m$, and final stellar mass, $m_f$, with more massive stars having higher accretion rates than lower mass stars: 
\beq 
\dot m = \dot m_{\rm TC} \left( \frac{m}{m_{f,1}} \right)^{1/2} m_{f,1}^{3/4},
\eeq
where $m_{f,1}\equiv \frac{m_f}{1M_\odot}$ and $\dot m_{\rm TC} \propto \Sigma_{c}^{3/4}$, the surface density of the core. However, this functional form describes a continuously increasing accretion rate for all times, which is inaccurate at late times when the core gas is depleted. To account for declining accretion, \citet{mckee10} proposed a tapering function $f_{\rm tap}=[1-t(m,m_f)/t_f(m_f)]$, where $t_f(m_f)$ is the formation time for a star with final mass $m_f$ such that $m(t_f) = m_f$. We find the accretion rate declines much more sharply than this tapering function and is better described by $f_{\rm tap}=[1-t(m,m_f)/t_f(m_f)]^4$. This corresponds to an accretion history given by:
\beq 
\dot m = \dot m_{\rm TC} \left( \frac{m}{m_{f,1}} \right)^{1/2} m_f^{3/4}\left[1-\left(\frac{m}{m_f} \right)^{1/2}\right]^2, \label{tcfit}
\eeq
 For this relation, accretion peaks when $m=1/9 m_f$ at a value of $\dot m = 4/27 \dot m_{\rm TC}m_{f,1}^{3/4}$. 
Although it is commonly assumed that accretion declines over the protostellar phase, the data are ambiguous on this point and  the exact functional form of the accretion rate is poorly constrained.  \citep{dunhamppvi14}. \citet{fischer17} found that the luminosities of young stellar objects in Orion could be explained by a model in which envelope masses exponentially decrease in time, a model that is consistent with the steep decline shown in Eqn.~\ref{tcfit}. 


Figure \ref{mdotvsmp} shows this tapered TC model provides a good description of the accretion histories. Both the location of the peak accretion and rate of decline agree well. The difference between curves occurs only due to the difference in final stellar mass. We set $m_f$ to be 10\% higher than the protostellar mass at the last simulation output since a small amount of accretion is continuing. Surprisingly, the magnetic field enters only indirectly through its influence on the final stellar mass.

\begin{figure}
\begin{center}
\includegraphics[scale=0.45]{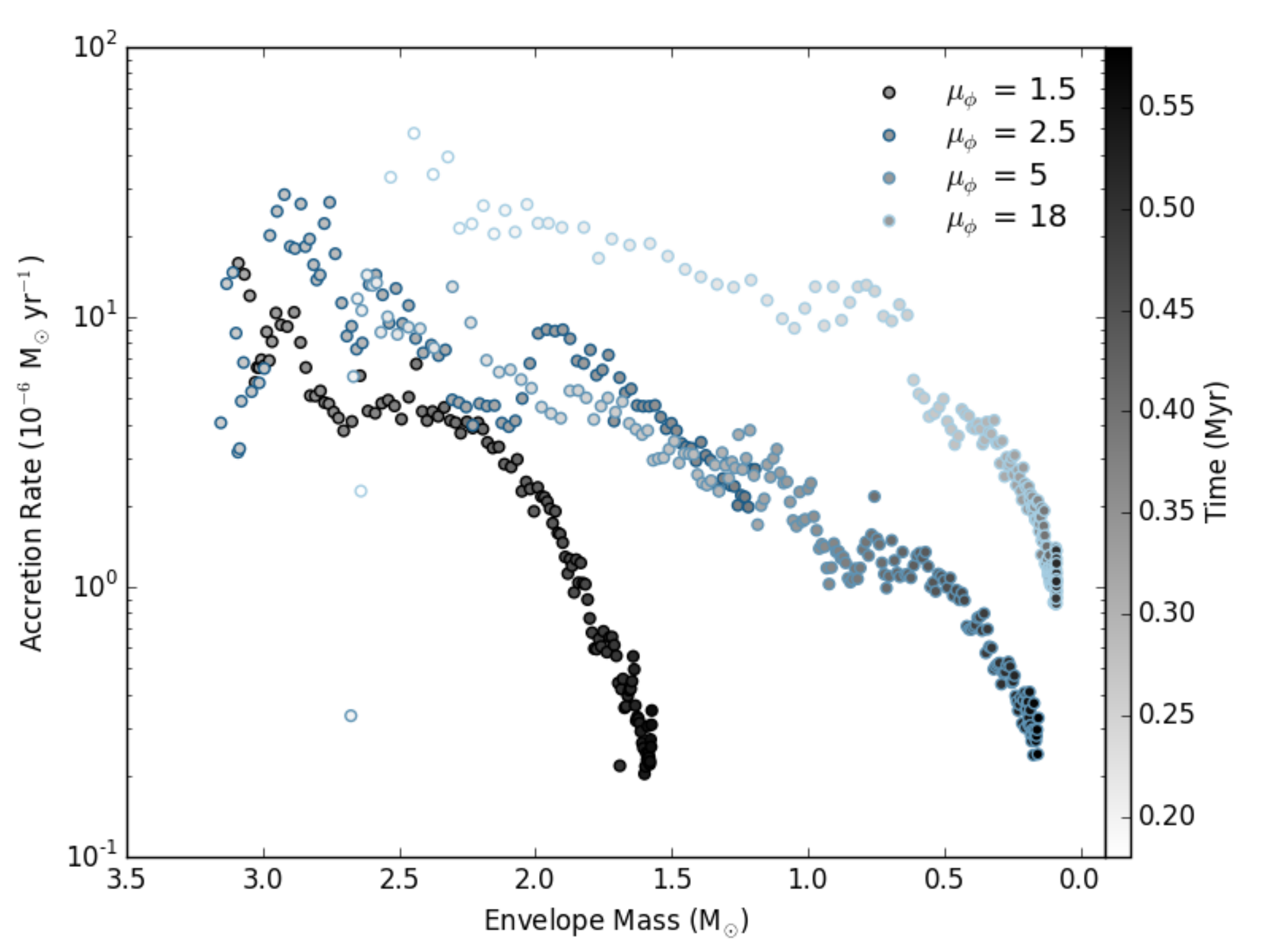}
\end{center}
\caption{ Averaged protostellar accretion rate as a function of envelope mass (gas with $n_{\rm H} \ge 10^4$ cm$^{-3}$). The colorbar indicates the time.
\label{mdotvsmenv} }
\end{figure}

Figure \ref{mdotvsmenv} shows the accretion rate as a function of the envelope mass, which is defined as gas with density $n_{\rm H} \geq 10^4$ cm$^{-3}$. During the main accretion phase for a given envelope mass the accretion rate varies over an order of magnitude. This suggests strong magnetic support can significantly regulate the accretion, and variation in  core magnetization alone may produce a factor of 10 spread in observed protostellar luminosities. Additional modulation of the accretion rate is produced by the core turbulence. However, we average over 32 consecutive fine time steps to eliminate luminosity variability due to small mass fluctuations in the accretion region. This corresponds to one time step on the basegrid or $\Delta t \sim 10$ years.

In the strongest field case, significant core gas remains even after accretion has declined by almost two orders of magnitude. This suggests the protostellar outflow is not sufficiently powerful to disperse the remaining core. For $\mu_\Phi\ge 2.5$, accretion remains sufficiently high for the outflow to disperse most of the core.

\begin{figure}
\begin{center}
\includegraphics[scale=0.45]{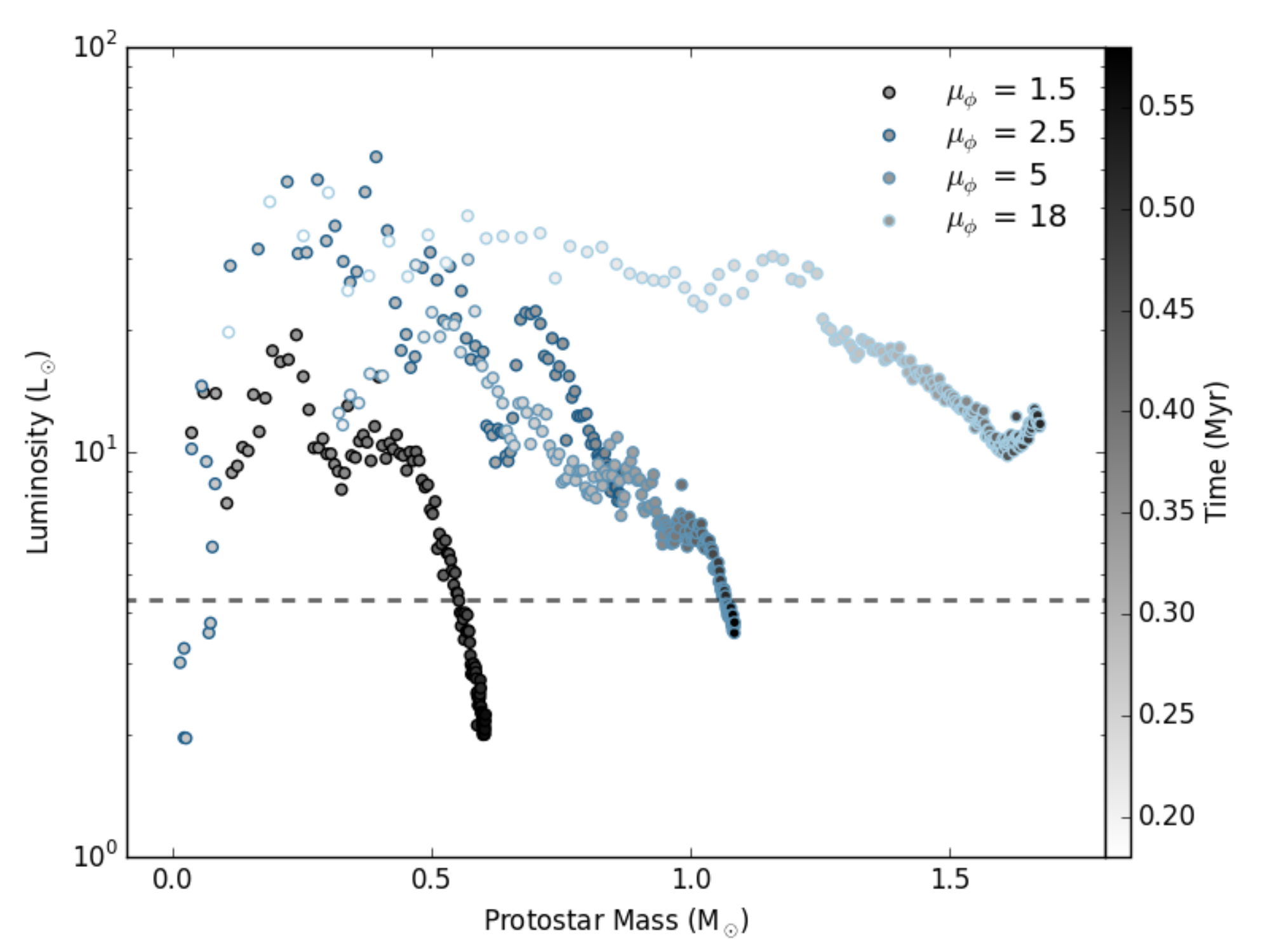}
\end{center}
\caption{ Protostellar luminosity as a function of protostar mass. The colorbar indicates the time. The dashed line indicates the mean bolometric luminosity, $\bar L = 4.3 L_\odot$, from objects in a combined c2d and Gould Belt survey by \citet{dunham13}.
\label{lvsmass} }
\end{figure} 

\begin{figure*}
\begin{center}
\includegraphics[scale=0.8]{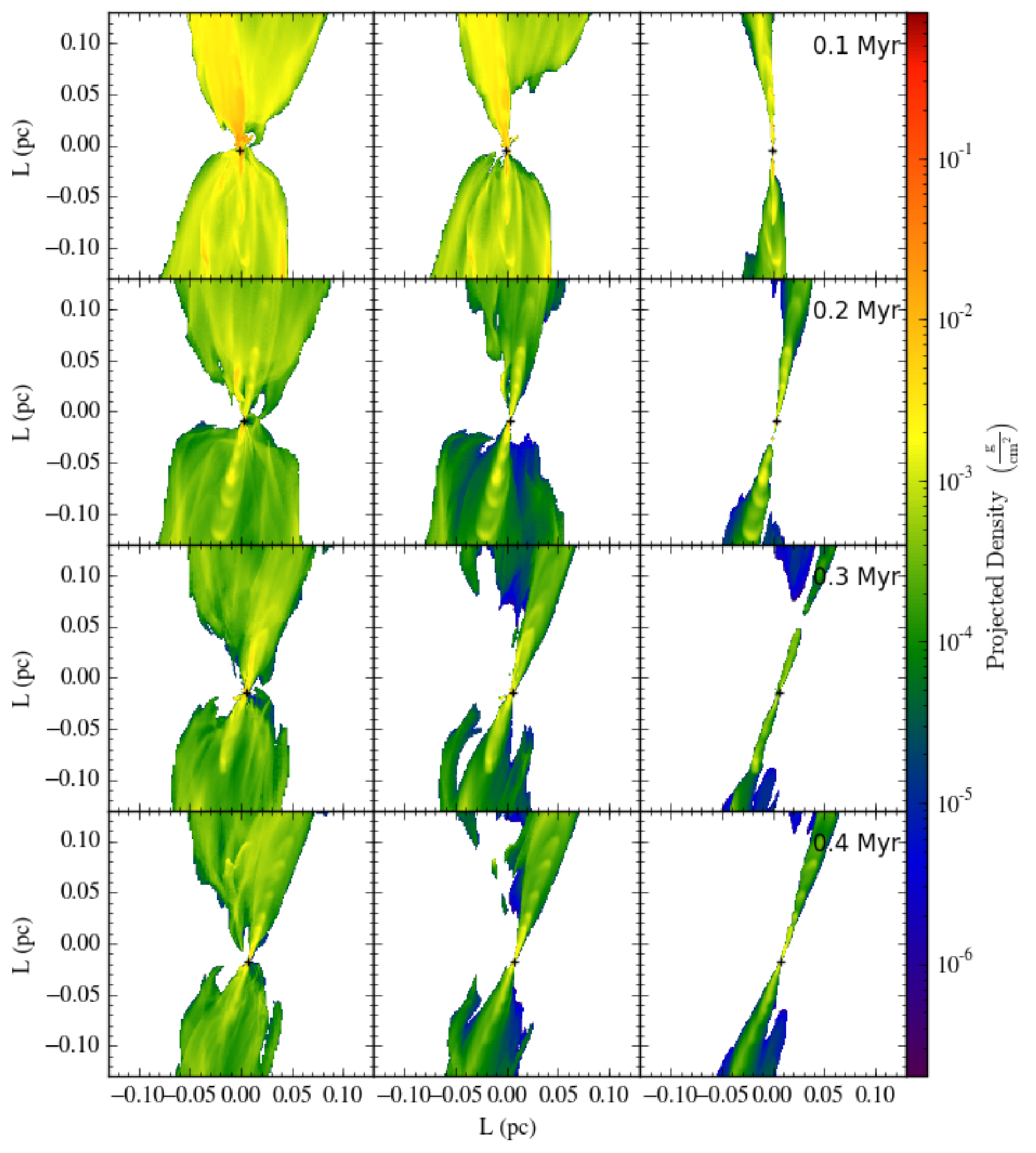}
\end{center}
\caption{Gas column density for gas contained in cells with $F_{\rm t} \geq$ 0.05, 0.10, and 0.25 (left to right) at four different times following protostar formation for run M4P18. The black crosses indicate the protostar location. \label{tracerfig}}
\end{figure*}

\subsubsection{Protostellar Luminosities}\label{lum}

For most times, accretion dominates the protostellar luminosity. The accretion luminosity is related to the protostellar mass and accretion rate by \citep[e.g.,][]{Offner11}:
\beq
L_{\rm acc} = f_{\rm acc} \space \frac{G m \dot{m }}{r},
\eeq
where $m$ is the protostellar mass, $\dot{m}$ is the accretion rate, $G$ is the gravitational constant, $r$ is the protostar radius, and the pre-factor $f_{\rm acc}$ = 0.75 is the accretion efficiency, describing the fraction of accretion energy that is radiated away \citep{shu94}. 
The similarity between the luminosity and accretion curves in Figure \ref{lvsmass} and Figure \ref{mdotvsmp}, respectively, underscores the direct relationship between the accretion rate and output luminosity.

The mean observed protostellar luminosity for distributions of protostars observed in nearby clouds ranges from 0.7 $L_\odot$-6.1$L_\odot$ \citep{dunhamppvi14}. Our simulated protostellar luminosities have time-averaged mean values of 5.5, 6.1, 7.6 and 11.5 $L_\odot$, for $\mu_\phi=1.5-18$,  respectively. The mean luminosity increases in proportion to the final stellar mass, which in all cases is above the expected mean mass for the observed samples of protostars.\footnote{ Assuming the stellar population will satisfy a Chabrier IMF \citep{chabrier05} and the highest mass star forming is $3 M_\odot$, the mean final mass is $\bar m_f \sim 0.4 M_\odot$.}

The luminosity trend suggests that younger protostars should have higher average luminosities than older protostars. 
This is tentatively supported by some comparisons of Class 0 and Class I objects \citep{fischer17}.
However, class, which is based on the spectral energy distribution, depends strongly on viewing angle and may not correlate well with actual source age \citep{young05,dunham12,Offner12}.

Our models suggest that luminosity variation for a given source can be driven by the magnetization of the core (factor of 10), variation of accretion with final mass (factor of 10-100), and turbulence (factor of a few). This may help to explain the spread in observed protostellar luminosities \citep[e.g.,][]{Offner11,dunhamppvi14}. Other proposed solutions include variable disk accretion and infall variation \citep{dunham12,padoan14}.



\begin{figure}
\begin{center}
\includegraphics[scale=0.62, trim={2.5cm 0 2.5cm 0},clip]{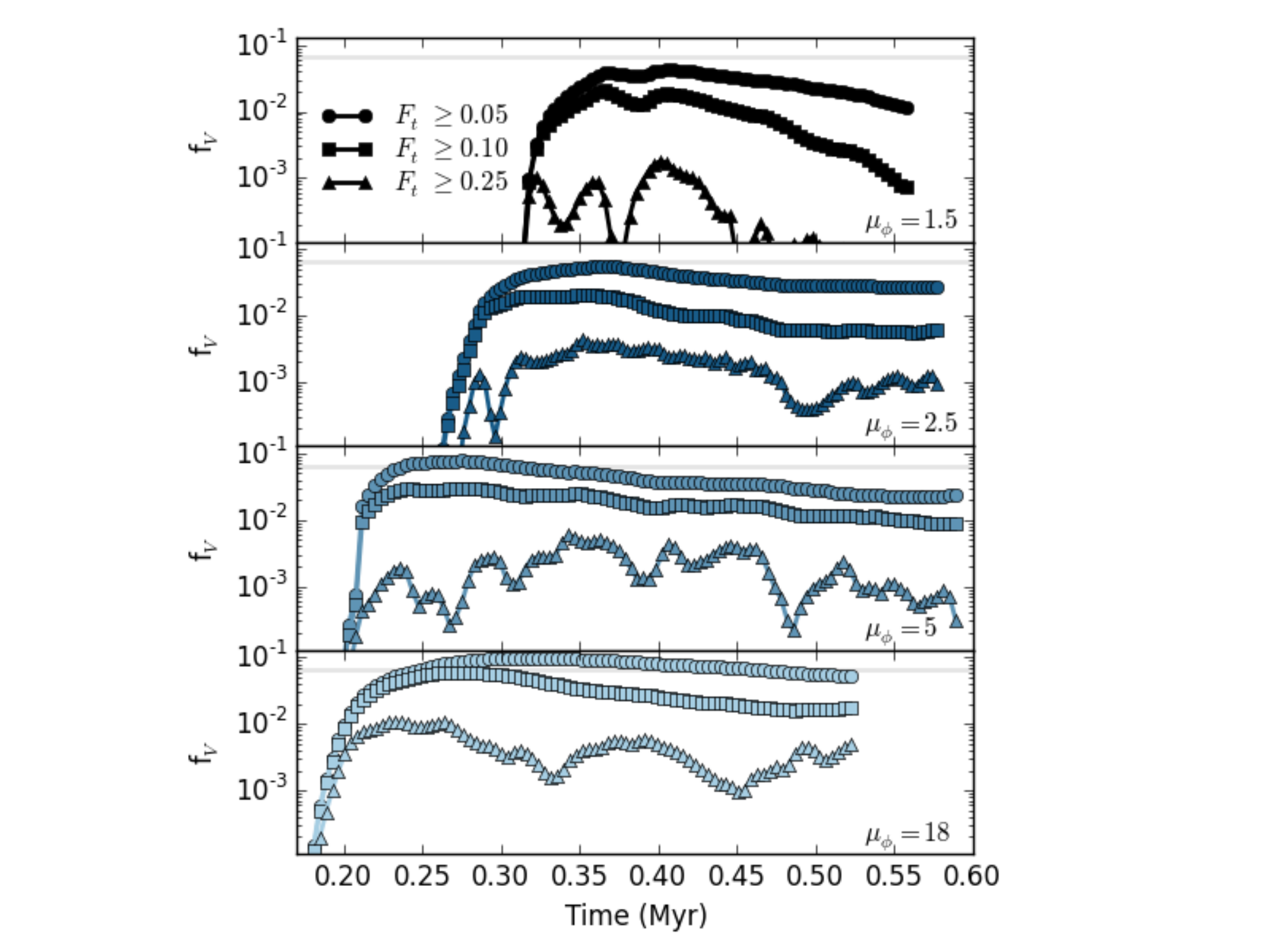}
\end{center}
\caption{ Volume fraction of the domain gas with $F_t \ge 0.05, 0.1, 0.25$ as a function of time for the four magnetized models. The solid gray line denotes the volume fraction of the initial dense core. 
\label{volfrac} }
\end{figure}

\subsection{Outflow Properties and Evolution}\label{outflowprop}

\subsubsection{Outflow Definition and Morphology}\label{morph}


By definition, the outflow is a collection of material accelerated away from the protostar. It is composed of two components: a high-velocity, well-collimated jet that originates close to the protostar and entrained core material that is swept along by the jet exodus or an associated disk wind. {\sc orion} tracks the location of the launched material using a tracer field, $\rho_t$, which specifies the amount of launched material in each cell on the domain.  
This tracer field presents a unique tool for examining the outflow morphology and efficiency of entrainment. 




Inspection of the tracer field shows that a large fraction of cells contain some amount of launched material, suggesting that significant mixing and entrainment occurs. Regions with large fractions of launched material have higher velocities and would be easily identified in observations as part of an outflow, whereas cells with small fractions and lower velocities would most likely be excluded from the identified outflow. In observations, some of this missing outflow gas can be recovered by introducing a model for the dense core  \citep{arce01, Offner11}.  

To examine the distribution of entrained material, we define the tracer fraction: 
\beq
F_{\rm t} =  \frac{\rho_{\rm t}}{\rho},
\eeq
where $\rho_{\rm t}$ is the tracer density and $\rho$ is the total gas density. This quantifies the relative presence of gas launched by the outflow in each cell. By projecting the cells with different tracer fractions along the coordinate axes we are able to observe the distribution of launched and entrained material throughout the protostellar core.

Figure \ref{tracerfig} shows the column density of cells with ratios $F_{\rm t} \geq$ 0.05, 0.10, and 0.25. The cells with $F_{\rm t} \geq 0.25$ (rightmost panels) are very collimated with an apparent opening angle of $\sim 5^\circ$. This cutoff produces a structure containing the highest velocity gas and resembles observed jets. In contrast, the definition $F_{t} \geq 0.05$ (leftmost panels) includes cells with lower gas velocities that exhibit much weaker collimation ($\sim 30^\circ-50^\circ$). This morphology resembles observed wide-angle molecular outflows, although some of these cells would probably not be included by visual identification. The mean velocity of this gas is much lower, in part because this threshold includes some gravitationally bound gas. The amount of launched material that exists in cells with $F_{\rm t} <$ 0.05 is a negligible fraction of the total launched material.

The cutoff $F_{t} \geq 0.1$ selects gas that appears moderately collimated with opening angles $10^\circ-40^\circ$. The different morphologies for different thresholds illustrate that a single launching mechanism can produce both a molecular and jet component, e.g., as proposed in early work by \citet{raga93}. 
For the remainder of our analysis, we adopt $F_{\rm t} \geq 0.1$ as the criterion to define gas in the outflow. We note this definition is independent of the gas velocity. In Appendix B we show that $F_{t} \geq 0.1$ also selects gas that is outflowing and unbound.

Figure \ref{tracerfig} highlights two secondary outflow behaviors. First, the outflow direction changes in time. This is caused by the accretion of turbulent material with varying net angular momentum, which impacts the launching direction.  As the direction changes, a relic outflow may be visible as in $F_{\rm t} \geq 0.05$ (bottom right panels in Fig.~\ref{tracerfig}). These lobes are no longer sustained by the jet, but the previously launched gas continues to move outwards, appearing as an apparently wider molecular outflow (leftmost panels). 
The outflow direction changes the most significantly during the high accretion phase ($<0.2$ Myr) and the direction stabilizes thereafter  \citep[see also][]{Offner16,lee17}. We find that the other models exhibit similar behavior. Evidence exists for outflow directional changes that are not clearly caused by precession \citep{kwon15,hsieh16,tan16}, but no systematic search has been carried out to date.

Second, close inspection of Figure \ref{tracerfig} shows small knots along the jet axis. These arise from nonuniform accretion (e.g., $t=0.2$ Myr panels).
Such knots are common in observations and are thought to result from episodic accretion \citep[e.g.,][]{dunhamppvi14,plunkett15,zhang16}. Here, they are produced by variable accretion even in the absence of a disk. The shape of the jet is also influenced by the interaction with the core gas, so the mass and velocity structure is not always symmetric \citep[see also][]{Offner11}. All the models exhibit this behavior independent of magnetic field strength. 


\subsubsection{Entrained Gas Volume}\label{entrainedv}

We define the volume fraction of entrained gas as the volume of cells with $F_{\rm t} \geq f_0$ normalized to the domain volume: $f_{V} = V_{\rm out, f_0}/L^3$. Figure \ref{volfrac} shows the volume fraction for tracer thresholds $f_0=0.05, 0.1$ and $0.25$. 

The lowest threshold case, $f_0=0.05$, includes material that is still gravitationally bound, and thus, serves as a proxy for the amount of turbulent mixing produced by the outflow. During the peak accretion phase,  Figure \ref{volfrac} shows the launched gas spreads throughout 5-10\% of the domain for all field strengths. This is sizable compared to the initial dense core volume of 6.5\% (gray solid line). The weakest field simulation, which forms the highest mass protostar, displays the most entrainment and turbulent mixing. In this case, the gas volume with $F_{\rm t} \geq  0.1$ also exceeds the initial core volume (see also Figure \ref{tracerfig}). 

In all runs, the cell volume for $F_{\rm t} \geq 0.25$ shows the most variability. These cells more directly trace the size and shape of the high-velocity jet, which can change considerably due to fluctuations in the accretion rate. Structure in these curves closely tracks variation in the accretion rate as shown in Figure \ref{mdotvsmp}. In contrast, the volume of gas for the two lower thresholds has a smooth  volumetric progression that tracks the average rather than instantaneous accretion.

\begin{figure}
\begin{center}
\includegraphics[scale=0.45]{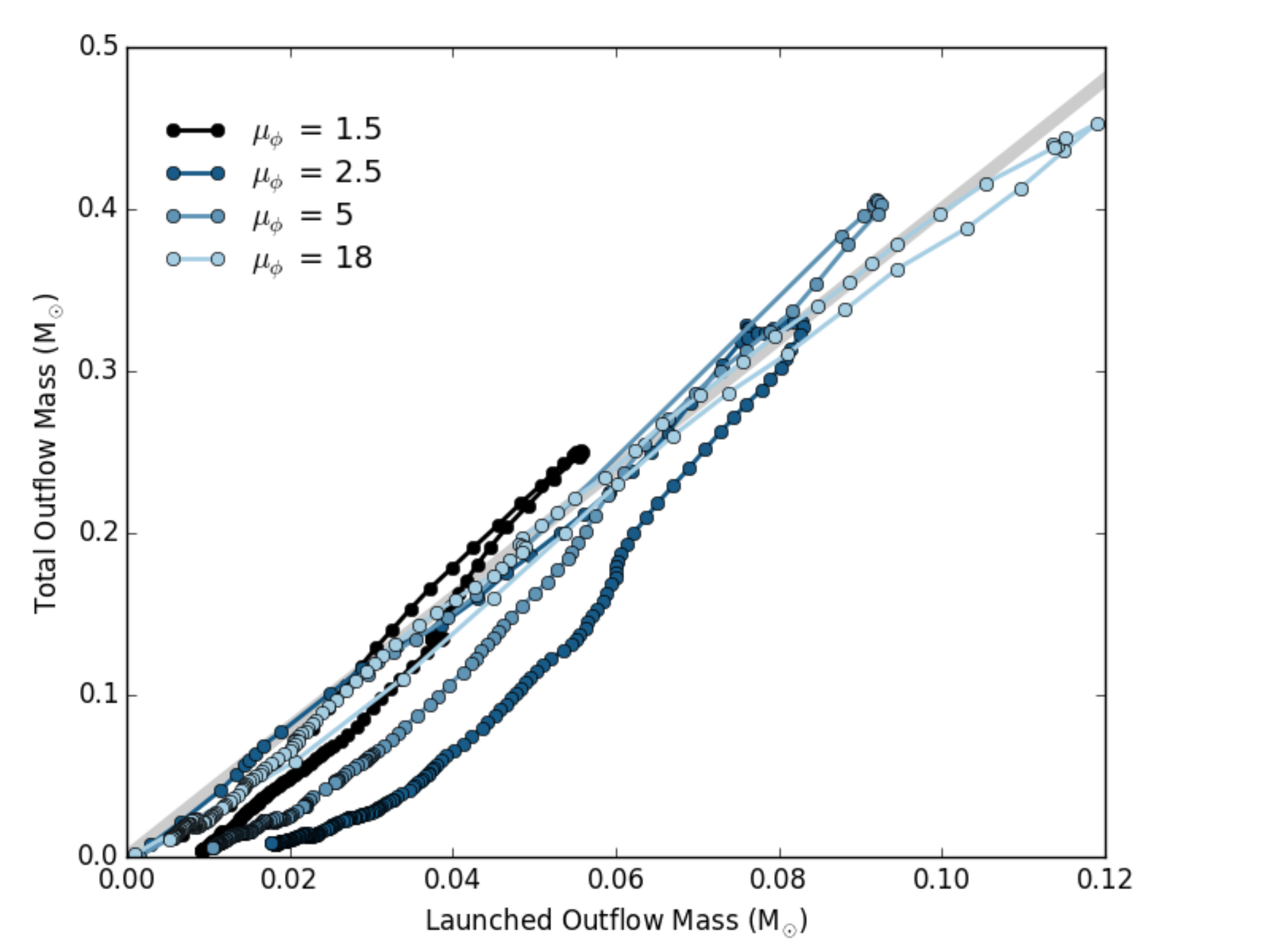}
\end{center}
\caption{ { Outflow mass versus launched mass for all times. The gray line has a slope of four ($M_{\rm out} = 4 M_{\rm launch}$). The black stars indicate the time of protostar formation. }
\label{entvslaunched} }
\end{figure}

\subsubsection{Entrained Gas Mass}\label{entrained}

We define the total outflow mass, $M_{\rm out}$, as the sum of all mass contained in cells with $F_{\rm t} \geq 0.1$. The middle panels in Figure \ref{tracerfig} show the column density of this gas for model M4P18. We define the launched mass, $M_{\rm launch}$, as the sum over the full tracer field. These quantities allow us to compare the amount of entrained and launched material over time. Figure \ref{entvslaunched} shows the outflow mass as a function of launched mass for all evolutionary times. 
It is striking that the outflow mass is about four times the launched gas mass for all times and magnetic field strengths. Consequently, at any given time the amount of entrained material is three times the amount of launched material, i.e., the entrainment efficiency is $\sim 300$\%. 

For all models, the outflow mass increases during the main accretion phase and then declines as accretion diminishes. At later times, the tracer material becomes more diffuse and the $F_t$ values in cells more distant from the main jet fall below the tracer threshold.
The maximum outflow mass is proportional to the mass-to-flux ratio and peak accretion rate. At this peak the total outflow mass is $\sim$ 10\% of the initial core mass. At later times, the amount of entrained material remains significant, comprising a few to 10\% of the dense gas mass  (e.g., see Fig. \ref{mvst}).

\begin{figure}
\begin{center}
\includegraphics[scale=0.45]{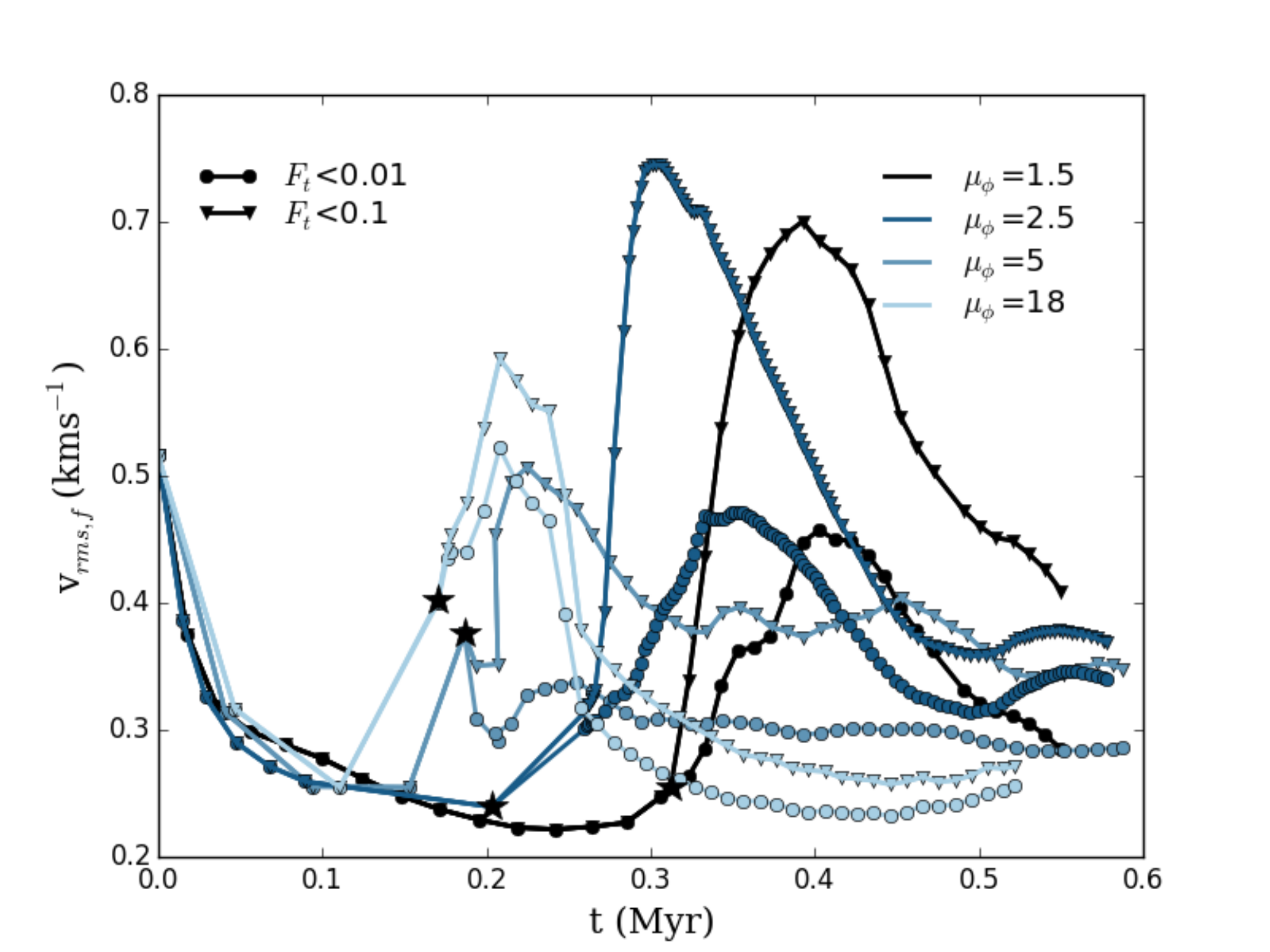}
\end{center}
\caption{ Mass-weighted velocity dispersion as a function of time for gas with $n_{\rm H}> 10^4$~cm$^{-3}$ and $F_t<0.1$ (circles) or $F_t<0.01$ (triangles).
\label{vfvst} }
\end{figure}


\subsubsection{Outflow-Driven Turbulence}\label{outdriving}

The tracer ratio provides a means to isolate the indirect impact of the outflow on the core gas, namely, non-local motions excited as a result of the core-outflow interaction. Figure \ref{vfvst} displays the velocity dispersion as a function of time for gas with $F_t <0.01$ and $F_t <0.1$. These cutoffs purposely exclude the main outflow and entrained material. Run M4P1.5 exhibits the largest velocity enhancement following the outflow launching, which suggests that the magnetic field magnifies the coupling between the outflow and core thereby increasing the effectiveness of outflow driving. These non-local motions are excited as the outflow ``plucks'' the magnetic field lines. The positive impact of the magnetic field on outflow-driven turbulence has previously been noted in simulations of clusters \citep[e.g.,][]{wang10}. The runs presented here illustrate the coupling between the outflow and field on core scales and the subsequent impact on the dense gas efficiency.  

The global velocity dispersion of M4P1.5 is the lowest of the runs as shown in Fig.~\ref{vvst}, and comparison of these two figures suggest that indirect outflow-driven turbulence accounts for most of the core kinetic energy at all times. Conversely, the weaker field runs exhibit progressively less non-local turbulence; the majority of the velocity dispersion during the main accretion phase is due to the outflow. Thereafter, the global velocity dispersion approaches $\sim$0.4-0.5 km/s, implying that at least half of the turbulence is produced indirectly by the outflow.

Figure \ref{vfvst} shows that the difference in velocity dispersion of gas with $F_t <0.01$ and $F_t <0.1$ increases with field strength. For the weakest field case, the velocity dispersion of gas with negligable launched material and a small amount of launched material is similar. This illustrates that magnetic coupling between the outflow and envelope can enhance momentum deposition in gas near the outflow ( $0.01 < F_t <0.1$ ), i.e., the momentum injected as a result of a small amount of mixing between the launched material and envelope deposits significantly more momentum as a function of field strength.

Removing the outflow from the velocity dispersion calculation has a secondary benefit. In observations, outflow material does not contribute to the linewidth by virtue of its density, temperature and chemistry. Figure \ref{vfvst} illustrates that the velocity dispersions of the dense gas excluding outflow material are in good agreement with observed 
values \citep{kauffmann13,kirk17}, after accounting for the $\sqrt{3}$ factor between our 3D dispersions and 1D observational measurements.

\begin{figure}[t]
\begin{center}
\includegraphics[scale=0.45]{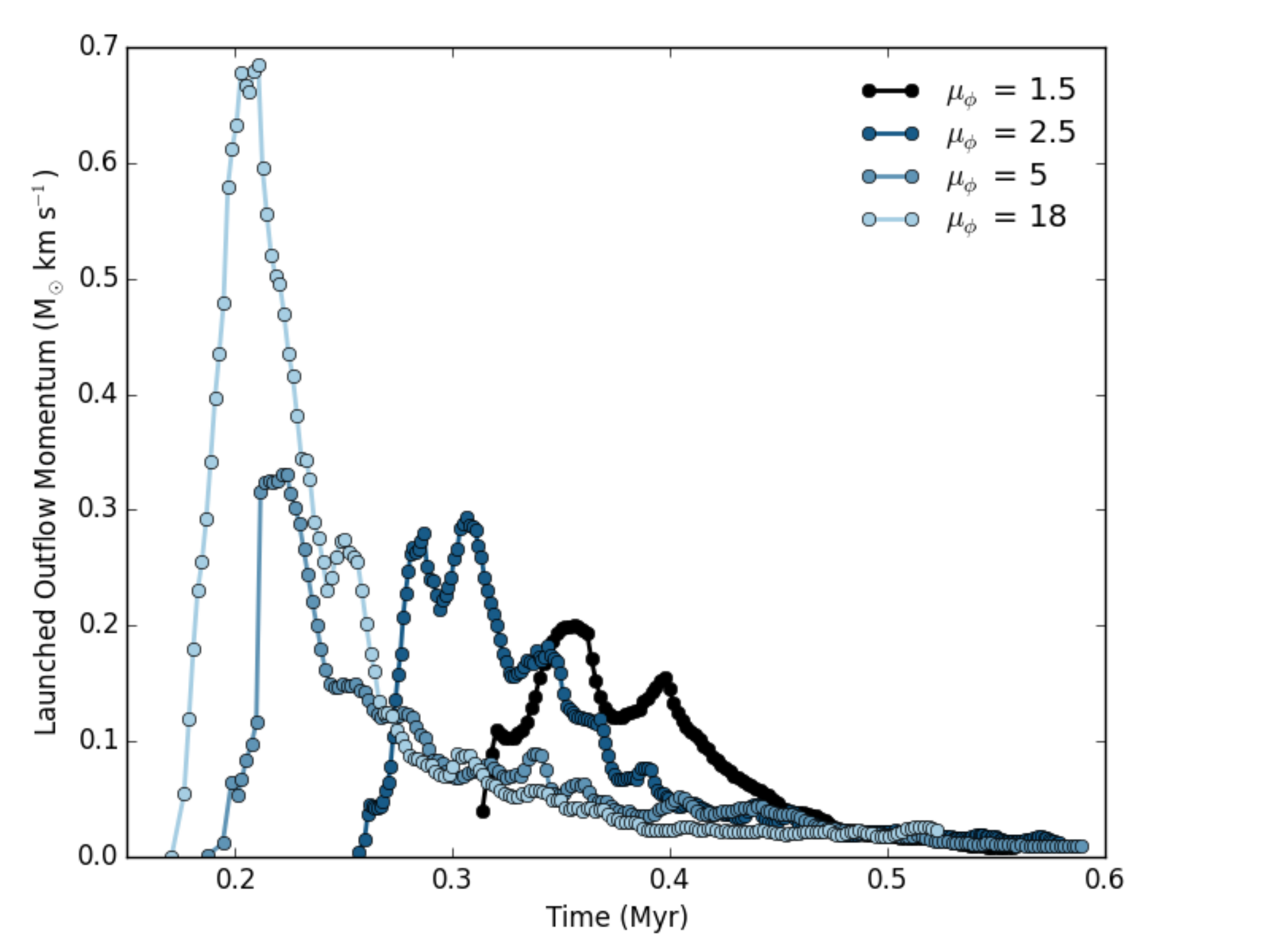}
\includegraphics[scale=0.45]{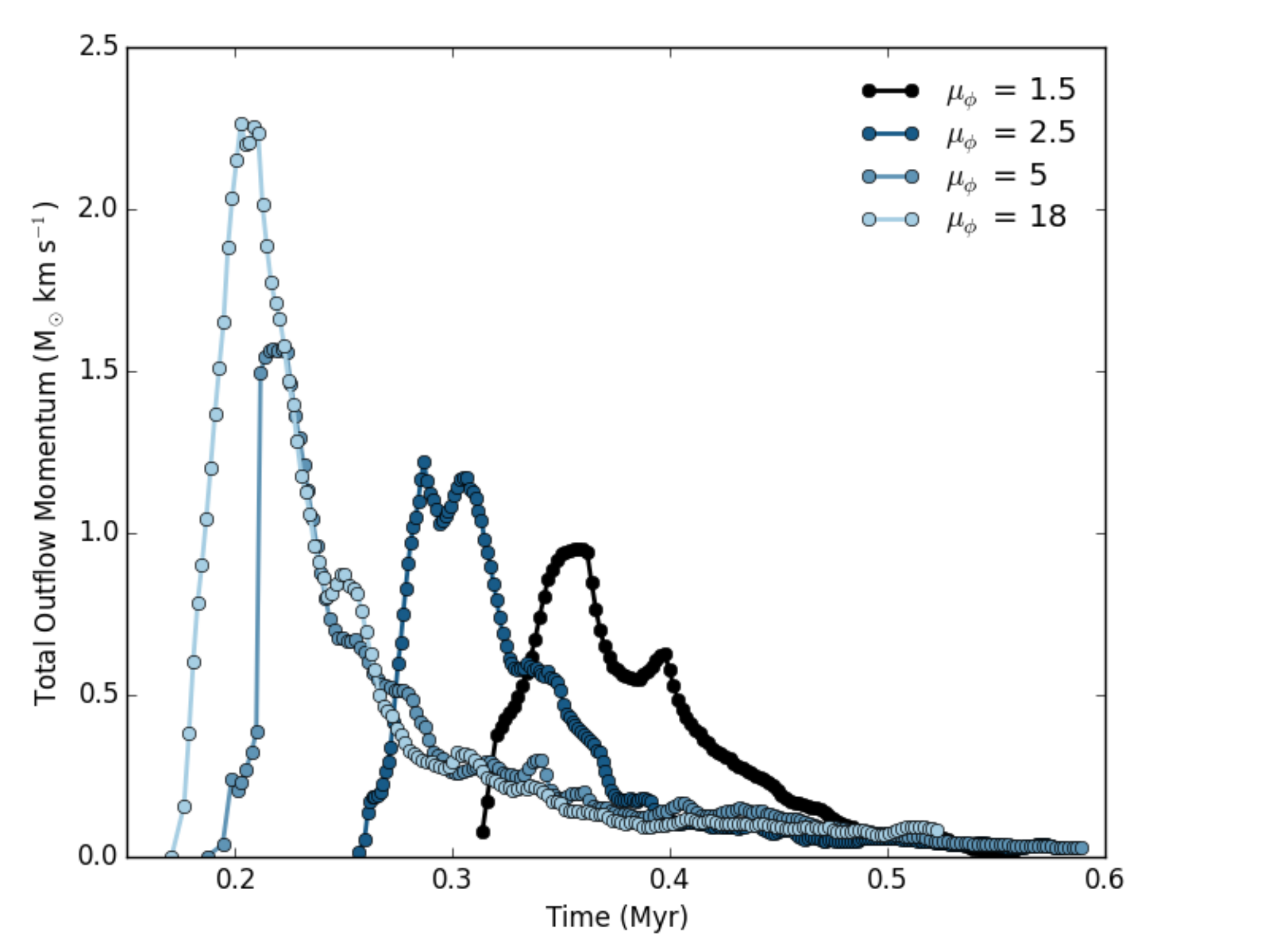}
\end{center}
\caption{ Top: Launched outflow momentum as a function of time. Bottom: Total outflow momentum (including entrained gas) as a function of time.
\label{momvst} }
\end{figure}

\subsubsection{Outflow Momentum}

By definition outflow feedback is momentum driven rather than energy driven, since radiative energy loss is efficient in molecular clouds \citep[e.g.,][]{krumholz14ppvi}. By comparing  momentum injection by outflows with the turbulent dissipation rate it is possible to determine whether turbulence can be rejuvenated by outflow feedback \citep{nakamura14}. Here we use the tracer field to investigate outflow momentum injection as a function of time and field strength.

We compute the momentum of the material launched by the outflow as
\beq
p_{\rm launch} = \Sigma \rho_{\rm t} v_{\rm rms} \Delta x^3 ,
\eeq
where $\rho_{\rm t}$ is the tracer field density, 
$v_{\rm rms}$ is the rms velocity, and the sum is computed over the full simulation domain. The total momentum of the outflow, $p_{\rm out}$, is computed similarly, replacing the tracer field with the gas density and summing over all cells with $F_{\rm t} \geq 0.1$.

Figure \ref{momvst} shows the launched and total momentum as a function of time for four models. The momentum evolution of both the launched and entrained material have similar features: they are dominated by high mass-loss at early times and by the increasing outflow velocity at late times. This transition occurs because the launched momentum is related to the protostellar mass and accretion rate by $p_{\rm launch} \propto \dot m v_{\rm K} \propto \dot m \sqrt{m}$.
The outflow momenta decline in tandem with the protostellar accretion, but  this trend is partially offset by the increasing protostellar mass and correspondingly higher launching velocity. 
Additionally, all curves demonstrate a local minimum $\sim 0.05$ Myr after protostar formation, which corresponds to the transition from mass- to velocity-dominated momentum. 

Several features differentiate the momenta as a function of initial magnetic field. The peak momentum injection changes dramatically between runs, with the lowest magnetic field strength having the highest momentum injection. This occurs because this run also experiences the highest accretion rate and hosts the most massive protstar. 


Once the outflow extent exceeds the domain size and outflow material begins to leave the simulation, the momentum values are lower limits. However, the figure shows that most of the momentum injection occurs during the first $\sim 0.1$ Myr.
Table \ref{starprop} lists the total momentum injected by the outflow model over the course of the simulation. \citet{arce10} find outflows in Perseus have mean momenta of $344\msun$ km s$^{-1}$/60 sources $\simeq 5.7~\msun$ km s$^{-1}$ after observational corrections. This is intermediate between the early inferred outflow momenta of $\sim 1-2~\msun$ km s$^{-1}$ in Figure \ref{momvst} and the total injected momentum in Table \ref{starprop}. This gives us confidence that our outflow model is qualitatively reproducing observed outflow characteristics. However, a more conclusive comparison with observations requires radiative transfer post-processing (e.g., OA14), which we reserve for future work. 

\section{Conclusions}\label{conclude}

We use magnetohydrodynamic simulations of isolated dense cores to explore the impact of magnetic field, initial core mass, turbulence and binarity on core evolution and star formation efficiency. This is one of the first studies of isolated low-mass cores to include turbulence, magnetic fields, radiative feedback from the protostar and a model for outflow launching. We track the material launched in the outflow using a special ``tracer field", which allows us to investigate outflow entrainment and momentum injection. We reach the following conclusions:
\begin{itemize}
\item Outflows drive and maintain transsonic turbulence in the protostellar environment even within highly magnetized cores. The strength of the initial magnetic field impacts the peak velocity dispersion, but the degree of turbulence at late times is largely independent of the field.
\item The time to protostar formation, and the subsequent Stage 0 lifetime, increases with the initial magnetic field strength. The Stage 0 lifetime ranges from $\sim 0.08$ Myr ($\mu_{\Phi}=18$) to greater than $0.35$ Myr ($\mu_{\Phi}=1.5$). These times are comparable to the observed Class 0 lifetime ($t_0=0.13-0.26$ Myr) in all but the strongest field case.
\item Magnetic fields increase core asymmetry and reduce accretion onto the protostar perpendicular to the field, leading to lower stellar masses.  A flattened central structure is produced, which may be prone to fragmentation on $\sim 1,000$ AU scales. 
\item Higher initial magnetic fields yield a lower star formation efficiency: $\epsilon \sim $0.4 ($\mu_{\Phi}=\infty$) - 0.15 ($\mu_{\Phi}=1.5$). These efficiencies can explain the observed offset between the core mass function and the stellar IMF 
without requiring high initial protostellar multiplicity. 
\item The protostellar accretion rate is well-described by a tapered turbulent core model, which is a function of the  fstellar mass. The magnetic field enters only indirectly through the mass of the protostar. Because of magnetic pressure support, the accretion rate, and hence protostellar luminosity, can vary by an order of magnitude for the same apparent envelope mass.
\item Gas visually identified as a highly-collimated protostellar jet has a launched gas fraction $\ge$ 25\%, i.e.,  as little as 25\% of jet mass is material that has been directly launched by the protostar, while the remainder is entrained gas. Gas visually associated with a wide-angle molecular outflow is composed of as little as 10\% launched material. 
\item For the visually-identified molecular outflow, the amount of entrained material relative to outflow material is {\it independent} of both the initial magnetic field strength and evolutionary time. The ratio of launched outflow mass to the total (combined launched and entrained) outflow gas mass is $\sim$ 1:4.
\item When initial turbulence is present, turbulent mixing driven by the outflow is relatively efficient, such that envelope gas with a launched gas fraction with $\ge$5\%  has a volume filling fraction of 5-10\% of the domain, which is comparable to the initial core volume. %
\item The peak momentum injection period spans $\sim$0.1 Myr, which coincides with the main accretion phase. The outflow momentum is anti-correlated with the initial magnetic field. For the 4 \msun core, the peak outflow momentum ranges from $\sim 1-2.3$ \msun km s$^{-1}$. 
\end{itemize}

These calculations lay the ground work for studies examining the origin of multiplicity in protostellar cores \citep{Offner16} and the interplay between magnetic fields and outflows \citep{lee17}. Future simulations are needed to investigate clustered star formation, in which core evolution may be influenced by interactions with the environment, to model core formation and to explore the impact of protostellar multiplicity on gas efficiency and turbulence.

\acknowledgements 
 Thanks to Christopher McKee, Chat Hull and an anonymous referee for comments that improved this manuscript.   
 SO acknowledges support from NSF grant AST-1510021.
The data analysis, images and animations were made possible by {\it yt} \citep{turk11}.
This work was supported in part by the facilities and staff of both the Yale University Faculty of Arts and Sciences High Performance Computing Center and by the Massachusetts Green High Performance Computing Center in Holyoke, MA.

\appendix
\section{A. Resolution Study}

\begin{figure}
\begin{center}
\includegraphics[scale=0.45]{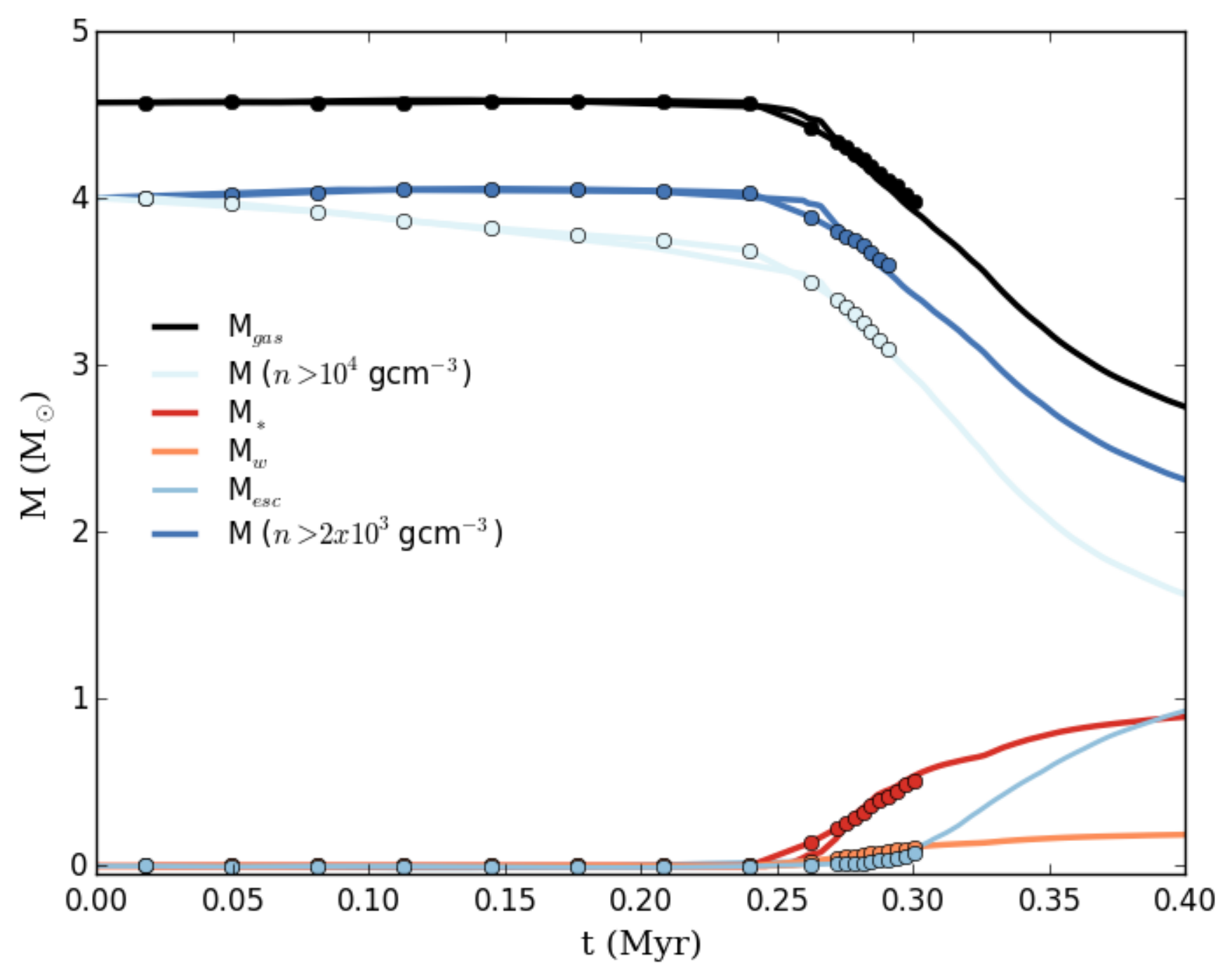}
\includegraphics[scale=0.45]{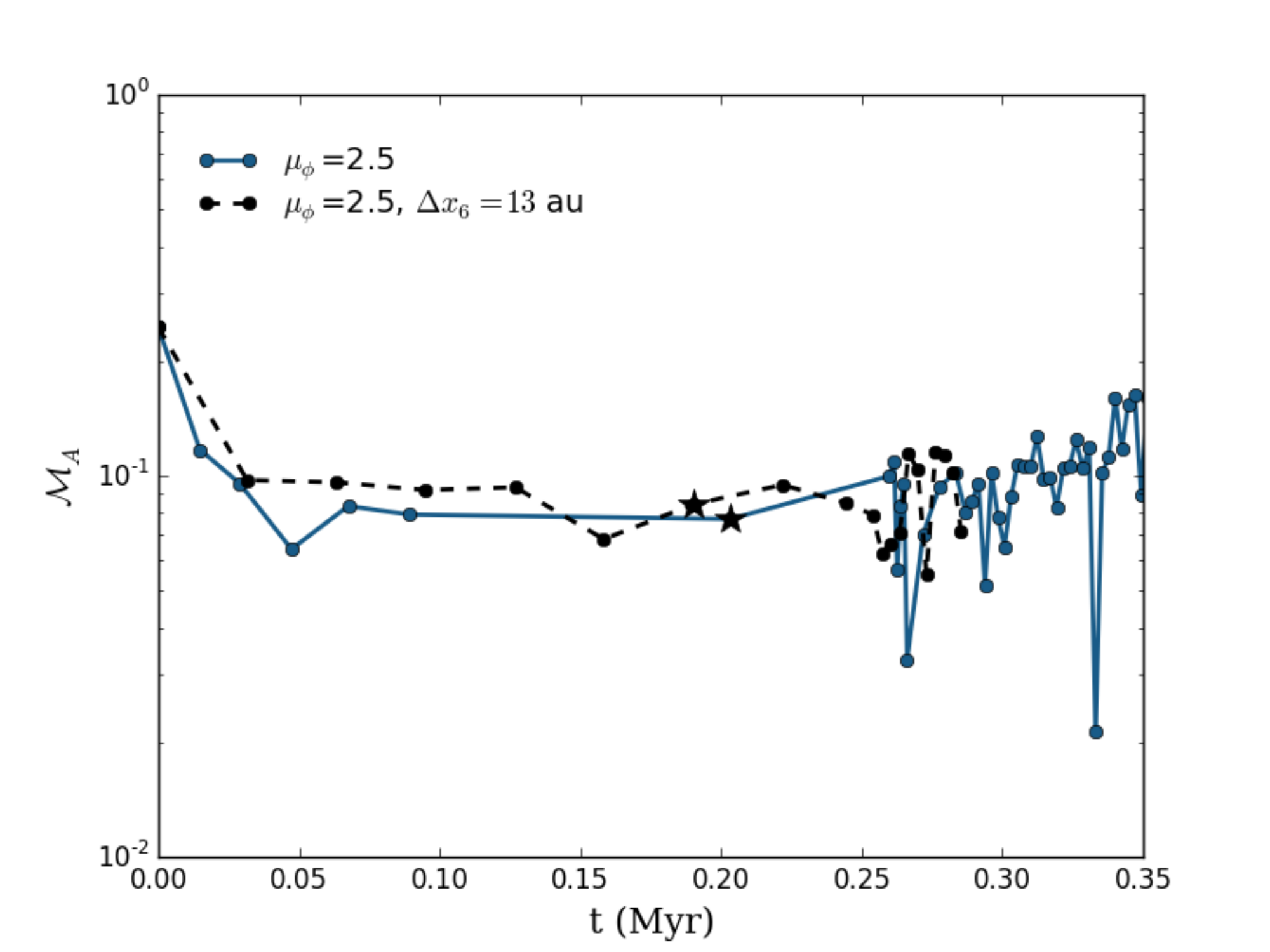}
\end{center}
\caption{ Left: Same as Figure \ref{mvst}. Time evolution of the total gas mass ($M_{\rm gas}$), core (dense) gas mass ($M_{\rm core}$), total mass launched in jet ($M_w$), protostellar mass ($M_*$), and mass escaped from the domain ($M_{\rm esc}$). The dotted lines show run the results for M4P2.5l6. The protostar forms slightly earlier in the higher resolution run, so the curves are offset by $\Delta t_{\rm o}=0.018$ Myr.  Right: Same as Figure \ref{mavst}. Alf\'ven Mach number as a function of time, where the dashed black line represents the result from the higher resolution run.
\label{mvstlev6} }
\end{figure}

In order to check the convergence of the calculations we re-run model M4P2.5 with an additional AMR level (M4P2.5l6). Figure \ref{mvstlev6} shows a comparison of the gas, protostar and outflow masses as well as the Alf\'ven Mach number for these two calculations. In the higher resolution run, the protostar forms slightly earlier, but thereafter, the two models show good correspondence. We are not able to follow model M4P2.5l6 for the same evolutionary time due to the prohibitive computational expense. See \citet{fielding15} for a more thorough resolution study of protostellar properties at different resolutions. 

We note that neither calculation has sufficient resolution to accurately resolve protostellar accretion disks. In the presence of magnetic fields, accretion disks are smaller and grid simulations require $\Delta x \sim$ au to resolve disk formation and follow the radial extent of the disk \citep[e.g.,][]{myers13}. No current MHD star formation calculations have sufficient resolution to accurately resolve three-dimensional disk structure including non-ideal effects and the magneto-rotational instability, which are thought to be fundamental to accretion and disk stability \citep{zhu10,zhu15}. 

\section{B. Outflow Definition and Gas Velocity}

The outflow definition introduced in \S\ref{outflowprop} does not take the gas velocity into account. In principle, some of the outflow material under our definition may not actually be outflowing, i.e., the gas radial velocity has $v_r < 0$ km s$^{-1}$ and is falling towards the protostar. Here, we verify that most of the gas with $F_t \geq 0.1$ also has positive radial velocity. In addition, we compare the outflow velocity to the escape velocity at edge of the initial dense core $v_{\rm esc} = \sqrt{2 GM_{\rm core}/r_{\rm core}} \simeq 0.7$ km s$^{-1}$ to estimate the fraction of outflow gas that is unbound.

\begin{figure}
\begin{center}
\includegraphics[scale=0.45]{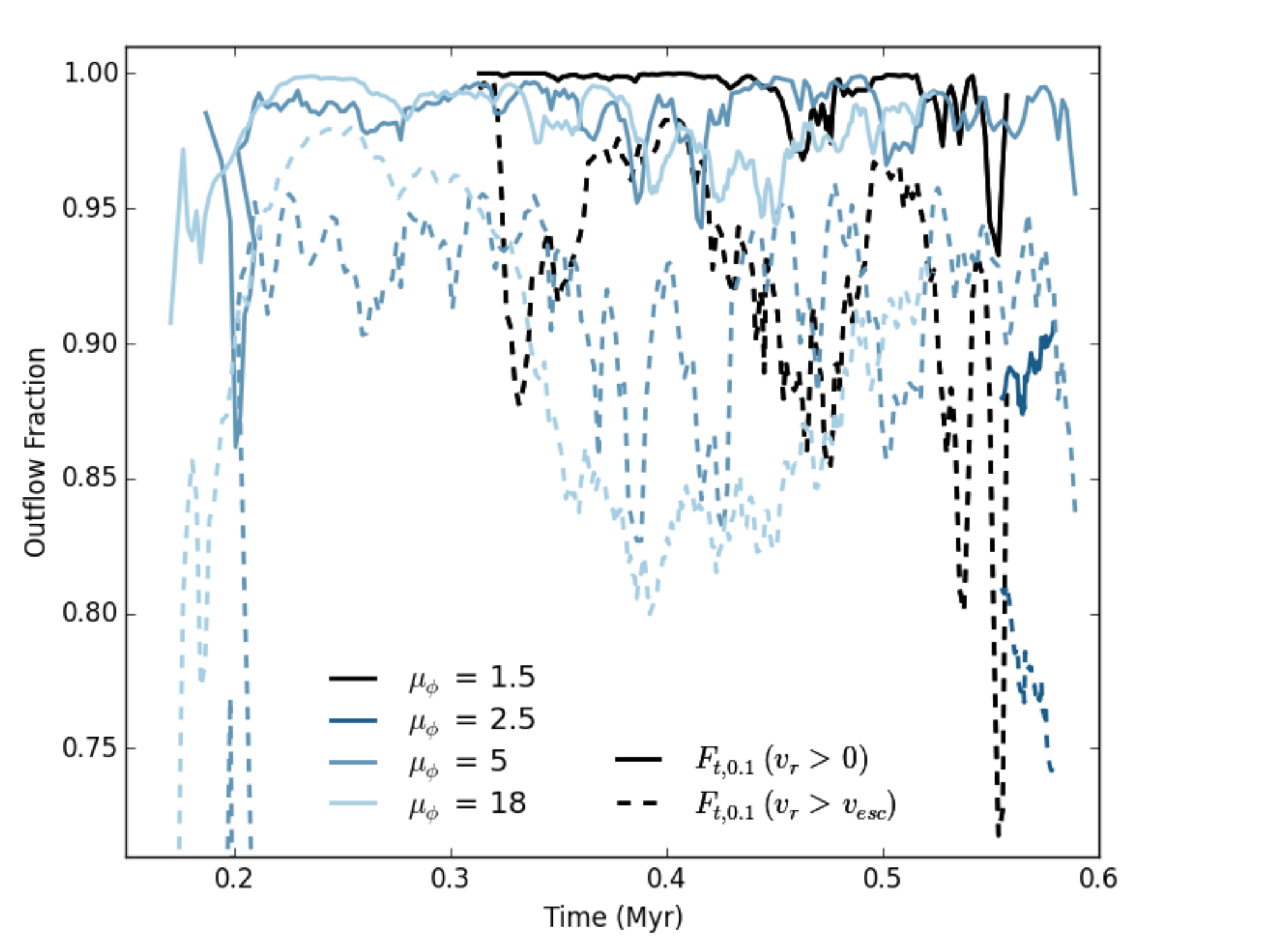}
\end{center}
\caption{ Fraction of entrained material -- as defined by $F_t \geq 0.1$ -- versus time for different magnetic field strengths. The solid curve indicates the fraction of cells with $v_r \geq 0$km s$^{-1}$. The dashed curve indicates the fraction of cells that have $v_r \geq v_{\rm esc} \sim 0.7$ km s$^{-1}$.
\label{tracervel} }
\end{figure}

Figure \ref{tracervel} shows that the $F_t$ parameter reflects the radial velocity quite well. On average, around 97\% of the gas included in the outflow has a positive radial velocity. There is a larger discrepancy when comparing the radial velocity to the escape velocity: about 10\% of the cells have velocities below $v_{\rm esc}$. Most commonly, the larger discrepancies appear later in simulation time. At these times the escape velocity is no longer well-approximated by the initial value, since significant mass has been expelled (e.g., Figure \ref{mvst}), 
lowering the local escape velocity for most of the gas of interest.   
In conclusion, this demonstrates that the choice of $F_t > 0.1$ effectively identifies truly outflowing gas.

\bibliography{outflowbib}
\bibliographystyle{apj}

\end{document}